\documentclass[twocolumn]{aastex631}
\usepackage{amsmath}
\usepackage{xcolor}
\usepackage{footmisc}
\usepackage{rotating,longtable}
\usepackage{tablefootnote}
\usepackage{booktabs}
\usepackage{comment}

\usepackage{graphicx}
\usepackage{natbib} 
\usepackage{hyperref}
\usepackage{tablefootnote}
\usepackage{booktabs}

\usepackage{lipsum,mwe}

\newcommand{\kms}{\,km\,s$^{-1}$}
\newcommand{\tco}{$^{13}$CO(1-0)}
\newcommand{\cetno}{C$^{18}$O(1-0)}
\newcommand{\co}{CO(1-0)}

\received{2023 December 7}
\accepted{2023 December 29}

\shorttitle{NGC2316}
\shortauthors{Sharma et. al.}

\begin{document}

\title{Cluster Formation in a Filamentary Cloud: The Case of the Stellar Cluster NGC 2316}

\author[0000-0001-5731-3057]{Saurabh Sharma}
\affil{Aryabhatta Research Institute of Observational Sciences (ARIES),
Manora Peak, Nainital 263001, India}
\author[0000-0002-6586-936X]{Aayushi Verma}
\affil{Aryabhatta Research Institute of Observational Sciences (ARIES),
Manora Peak, Nainital 263001, India}
\affil{M.J.P.Rohilkhand University, Bareilly, Uttar Pradesh-243006, India}
\author[0000-0002-3873-6449]{Kshitiz Mallick}
\affil{Aryabhatta Research Institute of Observational Sciences (ARIES),
Manora Peak, Nainital 263001, India}
\author[0000-0001-6725-0483]{Lokesh K. Dewangan}
\affil{Physical Research Laboratory, Navrangpura, Ahmedabad - 380009, India}
\author{Harmeen Kaur}
\affil{Center of Advanced Study, Department of Physics DSB Campus, Kumaun University Nainital, 263001, India}
\author[0000-0002-6740-7425]{Ram Kesh Yadav}
\affil{National Astronomical Research Institute of Thailand (Public Organization), 260 Moo 4, T. Donkaew, A. Maerim, Chiangmai 50 180, Thailand}
\author[0000-0002-0151-2361]{Neelam Panwar}
\affil{Aryabhatta Research Institute of Observational Sciences (ARIES),
Manora Peak, Nainital 263001, India}
\author[0000-0001-9312-3816]{Devendra K. Ojha} 
\affil{Tata Institute of Fundamental Research (TIFR), Homi Bhabha Road, Colaba, Mumbai - 400005, India}
\author[0009-0008-8490-8601]{Tarak Chand}
\affil{Aryabhatta Research Institute of Observational Sciences (ARIES),
Manora Peak, Nainital 263001, India}
\affil{M.J.P.Rohilkhand University, Bareilly, Uttar Pradesh-243006, India}
\author{Mamta Agarwal}
\affil{Aryabhatta Research Institute of Observational Sciences (ARIES),
Manora Peak, Nainital 263001, India}




\begin{abstract}
{
We present a multi-wavelength analysis of the star cluster NGC 2316 and its surroundings. We estimated
the physical parameters of the NGC 2316 cluster, including its shape (elongated), size (R$_{cluster}$ = 0.4 pc), distance ($1.3\pm0.3$ kpc), and minimum reddening ($A_V$ = 1.55 mag). We discovered two massive stars (B2.0V-B1.5V, age $\sim$12 Myr) embedded ($A_V$ = 4 mag) within this cluster. The cluster region still forms young stars even though the most massive star was born $\sim$12 Myr ago. We also found evidence of positive feedback from these massive stars. We identified a cold gas/dust lane extending westward from the cluster. 
The western end of the dust lane seems to favor low-mass star formation, whereas the cluster’s end favors bit massive star formation, which seems to have started earlier than the western end. We found an elongated molecular cloud in this region, characterized by numerous filamentary structures. The morphology of the filaments, along with position-velocity (pv) maps, velocity dispersion maps, channel maps, etc., indicate a coalescence of filaments and a potential longitudinal flow of matter toward the cluster through the western end of the gas/dust lane. This entire region seems to be a Hub-filamentary system (HFS), in which the NGC 2316 cluster is probably the hub and the dark lane is the main filamentary structure. Being the gravity well of this HFS, star formation started first in the NGC 2316 region and went on to the other filamentary nodes.
}        
\end{abstract}



\section{Introduction} \label{sec1}

Understanding the formation process of massive stars has been one of the leading research goals of many current and  past studies \citep{2016A&A...592A..54A,2017ApJ...835..142T,2018NatAs...2..478M,2018ApJ...859..166F,2021PASJ...73S.405F}  
as they have a profound effect on the Galactic evolution. Since massive stars start affecting their natal environment soon after their formation, these processes are even more complex \citep{2011AJ....141..123A}.
Stellar collision  \citep{1998MNRAS.298...93B}, competitive accretion \citep{2001MNRAS.323..785B}, monolithic collapse of a dense cloud and cloud-cloud collision (CCC;  \citealt{2021PASJ...73S.405F, 1992PASJ...44..203H}) are considered in the literature as plausible scenarios to explain the formation process of massive stars. 
%
Hub-filamentary systems (HFSs) have recently emerged as an essential site to look for massive stars and clusters and explain their formation \citep{2012A&A...540L..11S,2019A&A...631A...3M,2022A&A...658A.114K,2023MNRAS.tmp.3249D,2023JApA...44...23D,2023ApJ...958...51D}.

The filamentary structures are pervasive in the interstellar medium \citep{2010A&A...518L.102A,2010PASP..122..314M}. The \textit{Herschel} Space Mission and the subsequent investigations have apprised the involvement of the majority of the dense ($A_V>$ 7 mag) molecular gas in star formation in the form of filamentary structures \citep{2010A&A...518L.102A,2019A&A...621A..42A,2023ASPC..534..153H,2023ASPC..534..233P}. Filaments manifest diversity in the kinematic properties and physical ranges (refer \citealt{2023ASPC..534..153H,2023ASPC..534..233P}), which points towards different dynamical evolution. Filaments are noticed as a complex network with a parsec-scale central `hub' having high-column densities and low aspect ratios \citep{2009ApJ...700.1609M}.

\begin{figure*}
        \centering
        \hbox{
        \hspace{-0.5cm}
        \includegraphics[width=0.65\textwidth, angle=0]{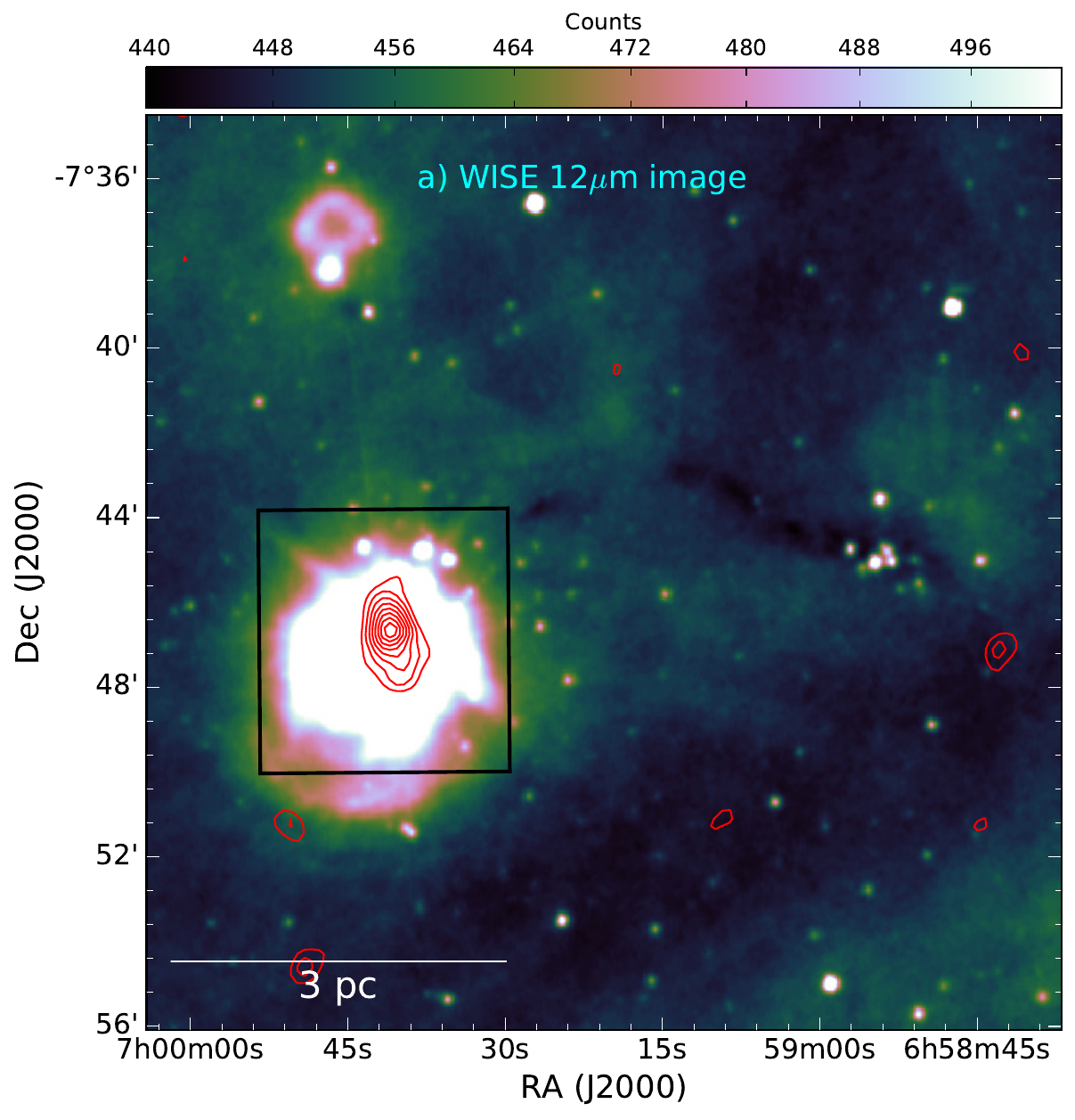}
        \hspace{-6.5cm}
        {\vbox
        {
        \includegraphics[width=0.3\textwidth, angle=0]{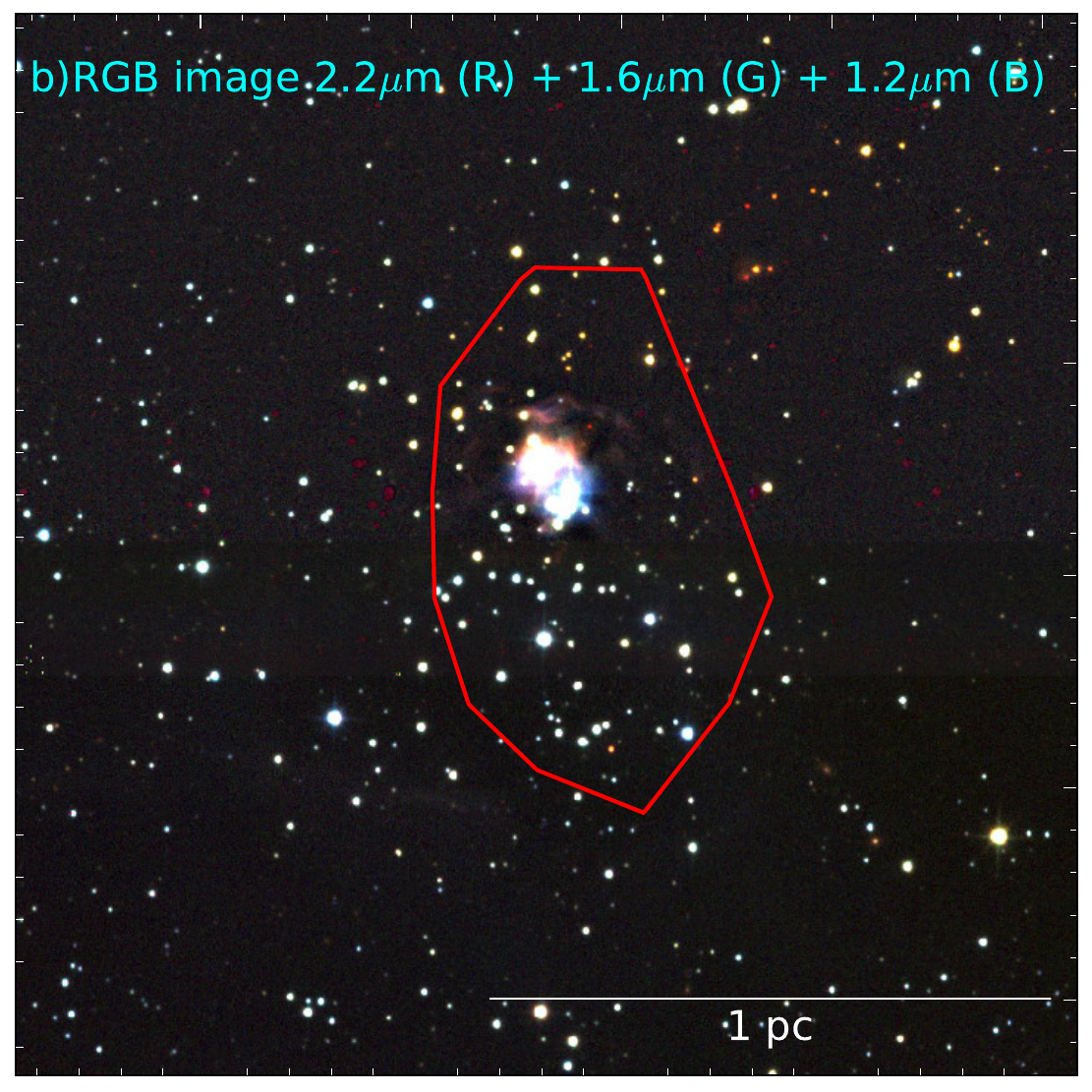}
        {\vbox{
        \vspace{0.8cm}
        \includegraphics[width=0.3\textwidth, angle=0]{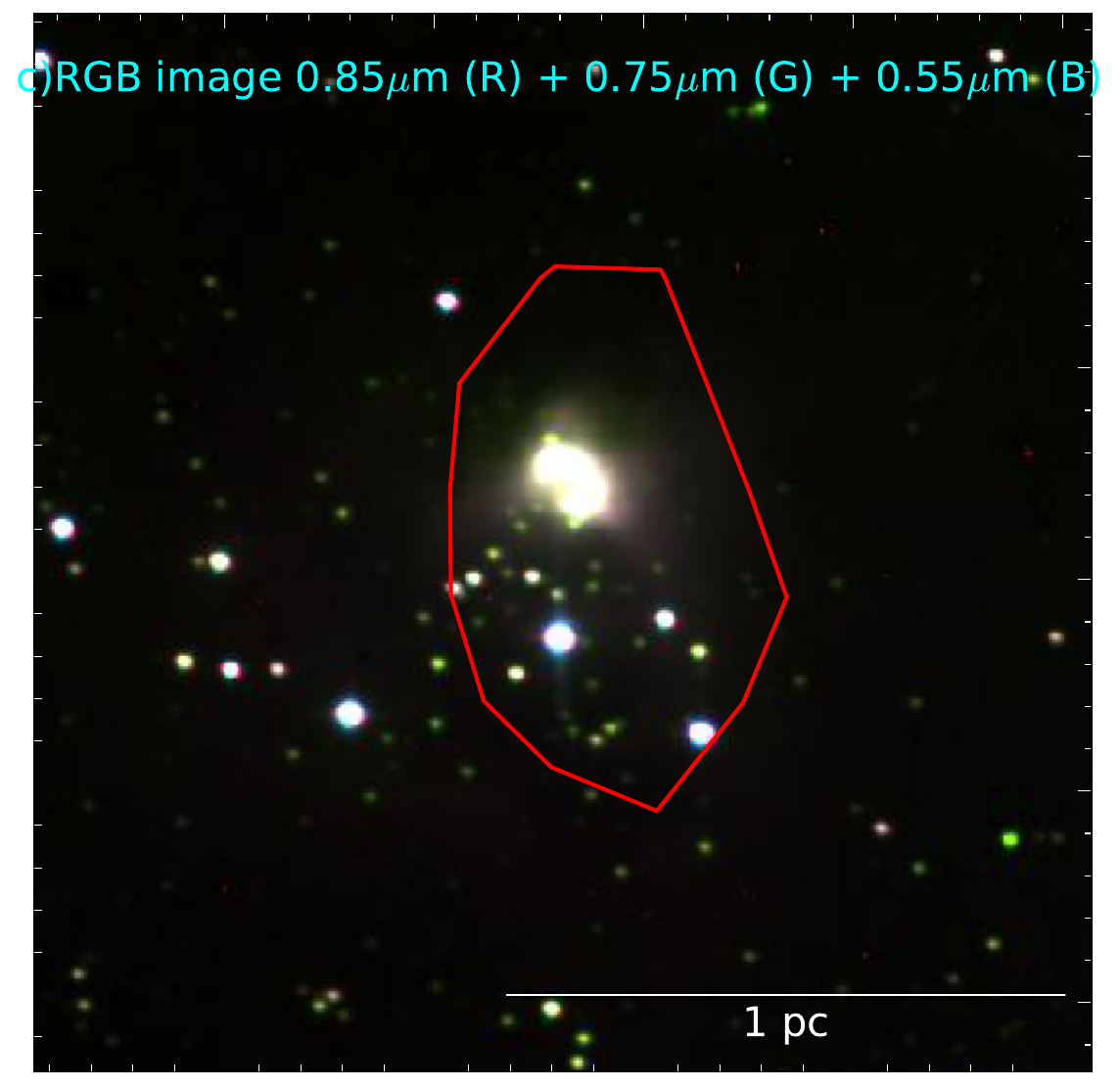}
        }}}}}
        \caption {  \label{3panel} Panel (a): WISE 12 $\mu$m MIR image for an area of $\sim20\times20$ arcmin$^2$ around the NGC 2316 cluster.
        The black square represents the FOV of the present observations using an optical imager on 3.6m DOT for the NGC 2316 cluster. The red contours are the isodensity contours generated using the nearest neighbor method from the near-infrared photometric data (cf. Section 3.1).
        Panels (b,c): Colour-composite images obtained by using the $J$ (blue), $H$ (green),
        and $K$ (red) images taken from UKIDSS (top panel), and the $V$ (blue), $R$ (green),
        and $I$ (red) images taken from the present optical observations (bottom panel). The red polygon is the \emph{convex hull} representing the cluster area generated on the outermost stellar isodensity contour (cf. Section \ref{group}).
        }
       \end{figure*}

NGC 2316 ($\alpha$$_{2000}$ =06$^{h}$59$^{m}$40$^{s}$, $\delta$$_{2000}$ = -07$\degr$46$\arcmin$36$\arcsec$) is a partially embedded young (2-3 Myr) star cluster located in an H\,{\sc ii} region
showing recent star formation activities, such as the distribution of photo-dissociated regions (PDRs; \citealt{1998MNRAS.294..338R,Ybarra_2014}), H$_2$ and Polycyclic Aromatic Hydrocarbon (PAH) emission \citep{Velusamy_2008}, signatures of ongoing star formation, such as H$_2$O masers \citep{1992A&A...255..293F}, and CO outflow with spectral index characteristic of optically thin free-free emission \citep{Beltrn2002IRAS2T}. \citet{Ybarra_2014} have identified shocked gas around the NGC 2316 cluster that may be part of an outflow originating from the cluster.
Figure \ref{3panel} shows the WISE 12 $\mu$m mid-infrared (MIR) image for an area of $\sim20\times20$ arcmin$^2$ around the NGC 2316 cluster.  The black square encloses the NGC 2316 cluster region. We can see the MIR emission in the region, which gives the warmed-up gas/dust distribution. This warming of the gas/dust usually happens due to the feedback from the massive stars. Also, we can see a dark dust lane from the cluster region (region showing very bright MIR emission) towards the western direction. At the western tip of the dark lane, we can see the distribution of a group of MIR point sources. These are all signatures of recent star formation activities with a hint of involvement of filamentary structures in the region. 

Thus, the NGC 2316 is an exciting region to study the star formation processes involving massive stars, young star clusters, molecular clouds, filamentary structures, and their impact on star formation.
In this paper, we have done a  multiwavelength study of this region, presented as follows. 
Section \ref{sec2} has information about the multiwavelength data sets used in this study and explains the data reduction procedures regarding our deep optical observations.
In Section \ref{sec3}, we analyzed the multiwavelength data sets to estimate the various morphological parameters and probed the physical environment 
around the NGC 2316 cluster. In section \ref{sec4}, we discussed the star-formation scenario in the NGC 2316 complex and summarized our study in Section \ref{sec5}.

\begin{table}
\centering
\caption{\label{log} Log of  observations.}
\begin{tabular}{@{}rrr@{}}
\hline
Date of  & Instrument/& Exp.$\times$No. of frames\\
observations& Filter   &(s)\\
\hline
2017 Dec 01&3.6m DOT/4K CCD &\\
$"$&$V$   &  $300\times11$\\
$"$&$I_c$   &  $300\times9$\\
2005 Jan 14 & 1.0m ST/2K CCD & \\
$"$&$U$   & $300\times3,1200\times3$ \\
$"$&$B$   &  $60\times3,600\times4$ \\
$"$&$V$   &  $60\times3,600\times4$ \\
$"$&$R_c$ & $60\times3,300\times4$\\
$"$&$I_c$ &   $60\times3,300\times4$\\
2005 Nov 04&$U$   & $300\times3$\\
$"$&$B$   &  $180\times3$\\
$"$&$V$   & $120\times1$ \\
$"$&$R_c$ &  $60\times3$ \\
$"$&$I_c$ &  $60\times3$\\
\hline

\end{tabular}
\end{table}

\section{Observation and data reduction}\label{sec2}

\subsection{Optical Photometric Observation and Data Reduction}

The deep optical photometric observations of the NGC 2316 cluster were carried out in $V, I_c$ bands on 2017 December 1, using 4K$\times$4K CCD IMAGER mounted at the axial port of 3.6m  Devasthal Optical Telescope (DOT), Nainital \citep{2018BSRSL..87...29K,2022JApA...43...27K}. The black square in the left panel of Figure \ref{3panel} represents $6^\prime.5 \times6^\prime.5$ field-of-view (FOV) of IMAGER on 3.6m DOT as a targeted region around the NGC 2316 cluster. We overlaid the image with stellar isodensity contours (please refer to Section 3.1) to highlight the NGC 2316 star cluster inside our targeted FOV. The right panel of Figure \ref{3panel} is a representation of the zoomed-in view of this region via optical and near-infrared (NIR) color-composite images overlaid with a \emph{convex hull} generated on the outermost stellar isodensity contour (please refer to Section \ref{group}).
The optical images from 3.6m DOT were taken in 2$\times$2 binning mode for a total integration time of 55 minutes and 45 minutes in $V$ and $I_c$ bands, respectively. 
We also observed the NGC 2316 cluster by 1.0m Sampurnand Telescope (ST), Nainital, in broadband $U$, $B$, $V$, $R_C$, and $I_c$ filters
using the 2K $\times$ 2K CCD camera having a FOV of $13\arcmin.5 \times13\arcmin.5$ \citep{2012ASInC...4..173S,2020ApJ...896...29K}. 
The complete log of observations of NGC 2316 is given in Table \ref{log}.

The basic data reduction, including image cleaning, photometry, and astrometry, was done using the standard procedure explained in \citet{2020MNRAS.498.2309S} and \citet{2020ApJ...896...29K}. 
NGC 2316 field was observed on the same night (2005 November 4) as NGC 6910 cluster \citep{2020ApJ...896...29K} along with a standard field \citep[SA98,][]{1992AJ....104..340L}. Thus, we used the same calibration equations, as \citet{2020ApJ...896...29K}, to transform instrumental magnitudes of several bright stars in the NGC 2316 field into standard Vega systems. These stars (local secondary standard stars) were then used to calibrate the instrumental magnitudes of all the stars detected in the NGC 2316 field through the 3.6m DOT observations.

\begin{figure}
    \centering
    \includegraphics[width=0.48\textwidth]{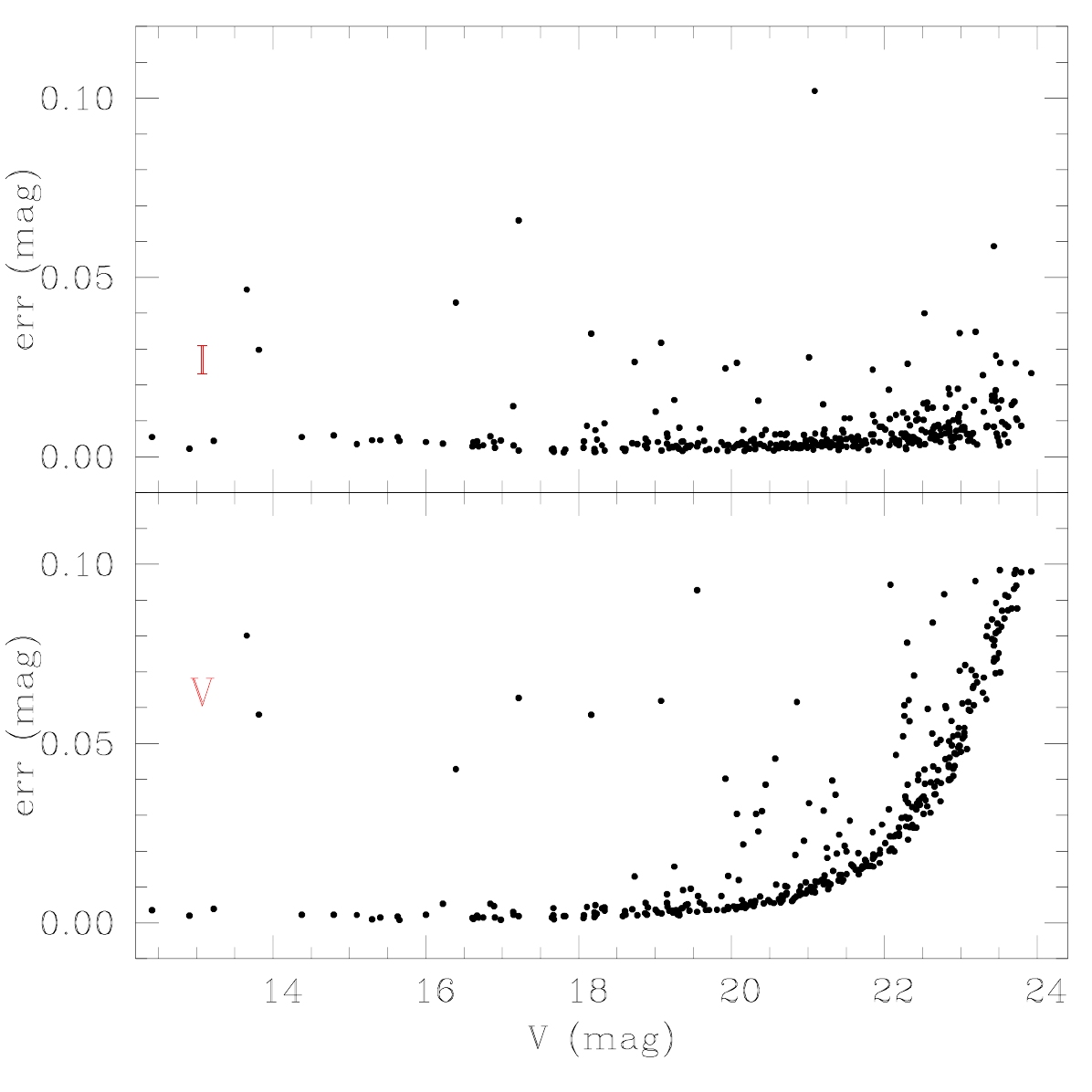}
    \caption{ Photometric error vs. magnitude plots in $V$, and $I_c$ filters. 
    \label{fig:optical_mag}}
\end{figure}

We compared the present standard magnitudes in $V$ and $B$ bands with archive `APASS'\footnote{The AAVSO Photometric All-Sky Survey, https://www.aavso.org/apass} and found no shifts in our calibrated magnitudes. 
The stars were identified with detection limits (photometric error $\leq0.1$, cf. Figure  \ref{fig:optical_mag}) of $\sim$24.0 in $V$ band.
We used Graphical Astronomy and Image Analysis Tool\footnote{http://star-www.dur.ac.uk/~pdraper/gaia/gaia.html} to do the astrometry of the stars with rms noise of the order of $\sim0\arcsec.3$.

\subsection{Molecular line data}
\label{section_MolecularCOData}

We obtained the archival \co. \tco, and \cetno, molecular line data of 1$^\circ$ $\times$ 1$^\circ$ region observed as a part of MWISP (Milky Way Imaging Scroll Painting) project by the PMO (Purple Mountain Observatory) 13.7\,m millimeter-wave radio telescope (\citealt{ 633694461037117445}; \citealt{Su_MWISP_2019ApJS}).
The channel width of these spectral cubes were $\sim$ 0.16 \kms, 0.17 \kms, and 0.17 \kms, respectively; with an rms noise of $\sim$\,0.45\,K, $\sim$\,0.22\,K, and $\sim$\,0.22\,K, respectively (the pixel brightness is in T$_{\mathrm{MB}}$ - main beam temperature). These cubes' spatial resolution and grid sizes are $\sim$\,50\arcsec\, and 30\arcsec, respectively. 

\subsection{Other Archival Data}

\begin{table*}[]
    \footnotesize
    \centering
    \caption{List of surveys adopted for the present work (NIR to Radio Wavelength).}
    \begin{tabular}{c c c c}
    \hline
    Survey & Wavelength/s & $\sim$ Resolution & Reference\\
    \hline
    Two Micron All Sky Survey\footnote{\citet{https://doi.org/10.26131/irsa2}} (2MASS) & 1.25, 1.65, and 2.17 $\mu$m & $2\arcsec.5$ & \citet{2006AJ....131.1163S}\\
    \emph{Gaia} DR3\footnote{https://www.cosmos.esa.int/web/gaia/dr3} (magnitudes, parallax, and PM) & 330–1050 nm & 0.4 mas & \citet{2016gaia,2023gaia}\\
    \emph{Herschel} Infrared Galactic Plane Survey\footnote{http://archives.esac.esa.int/hsa/whsa/} & 70, 160, 250, 350, 500 $\mu$m & $5\arcsec.8$, $12\arcsec$, $18\arcsec$, $25\arcsec$, $37\arcsec$ & \citet{2010PASP..122..314M}\\
    NRAO VLA Sky Survey\footnote{https://www.cv.nrao.edu/nvss/postage.shtml} (NVSS) & 21 cm & $46\arcsec$ & \citet{1998AJ....115.1693C}\\
    \emph{Spitzer} GLIMPSE360 Survey\footnote{\citet{https://doi.org/10.26131/irsa214}} & 3.6, 4.5 $\mu$m & $2\arcsec$, $2\arcsec$ & \citet{2005ApJ...630L.149B}\\
    Wide-field Infrared Survey Explorer\footnote{\citet{https://doi.org/10.26131/irsa1}} (\emph{WISE}) & 3.4, 4.6, 12, 22 $\mu$m & $6\arcsec.1$, $6\arcsec.4$, $6\arcsec.5$, $12\arcsec$ & \citet{2010AJ....140.1868W}\\
    UKIRT InfraRed Deep Sky Survey\footnote{http://wsa.roe.ac.uk/} (UKIDSS) & 1.25, 1.65, and 2.22 $\mu$m & $0\arcsec.8$, $0\arcsec.8$, $0\arcsec.8$ & \citet{2008MNRAS.391..136L}\\
    Milky Way Imaging Scroll Painting (MWISP)  & \co, \tco, \cetno  &  $50\arcsec$ & \citet[]{Su_MWISP_2019ApJS} \\
    \hline
    \end{tabular}
    \label{tab:archival_data}
\end{table*}

We used our target source's archival data sets ranging from NIR to radio wavelength regimes. Table \ref{tab:archival_data} briefly specifies these data sets. The \emph{Herschel} column density and temperature maps (spatial resolution $\sim12\arcsec$) were downloaded directly from the publicly available website\footnote{http://www.astro.cardiff.ac.uk/research/ViaLactea/}. These maps are procured for EU-funded ViaLactea project \citep{2010PASP..122..314M} adopting the Bayesian PPMAP technique \citep{2010A&A...518L.100M} at 70, 160, 250, 350, and 500 $\mu$m wavelengths \emph{Herschel} data \citep{2015MNRAS.454.4282M, 2017MNRAS.471.2730M}.

\begin{figure*}
\centering
\includegraphics[width=0.48\textwidth]{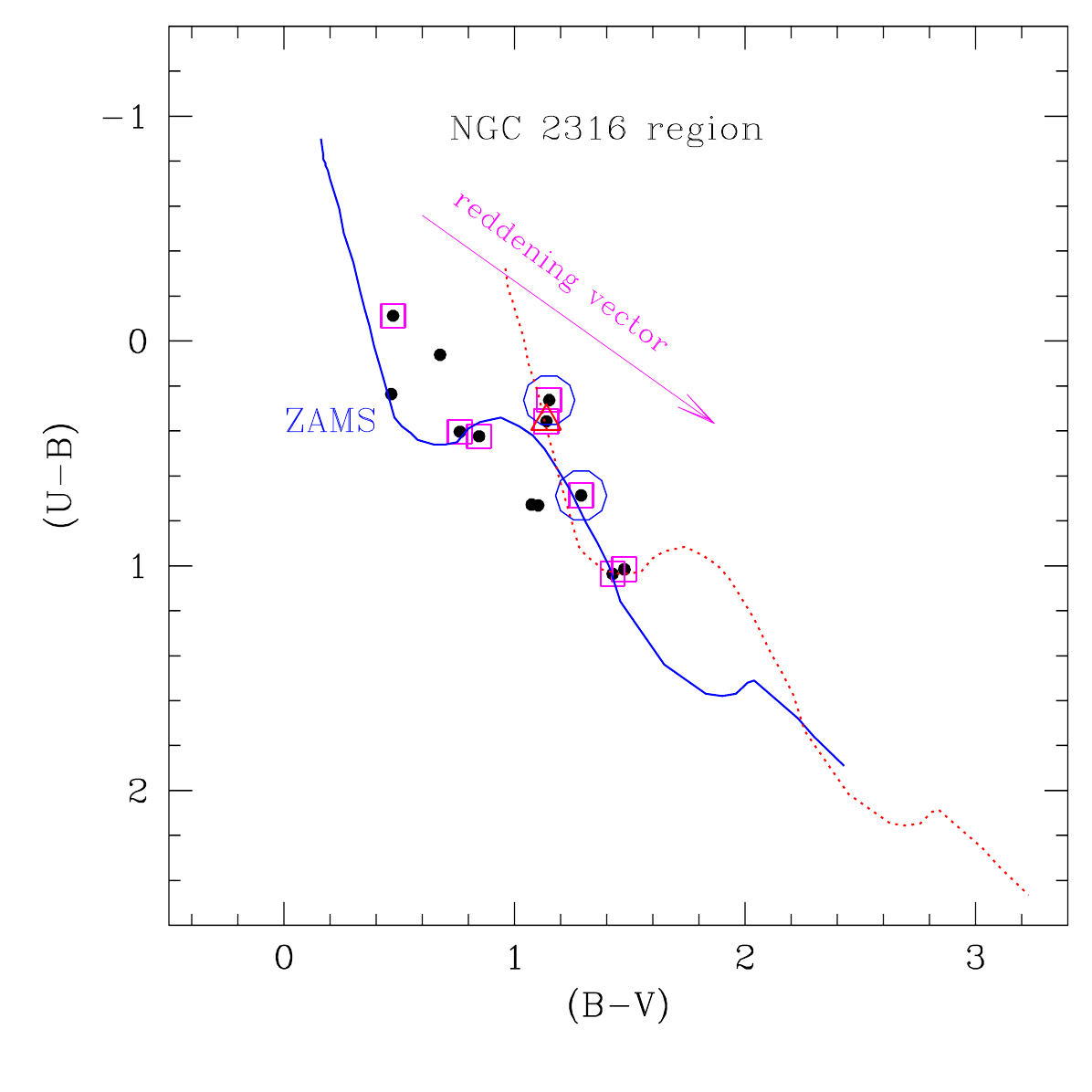}
\includegraphics[width=0.48\textwidth]{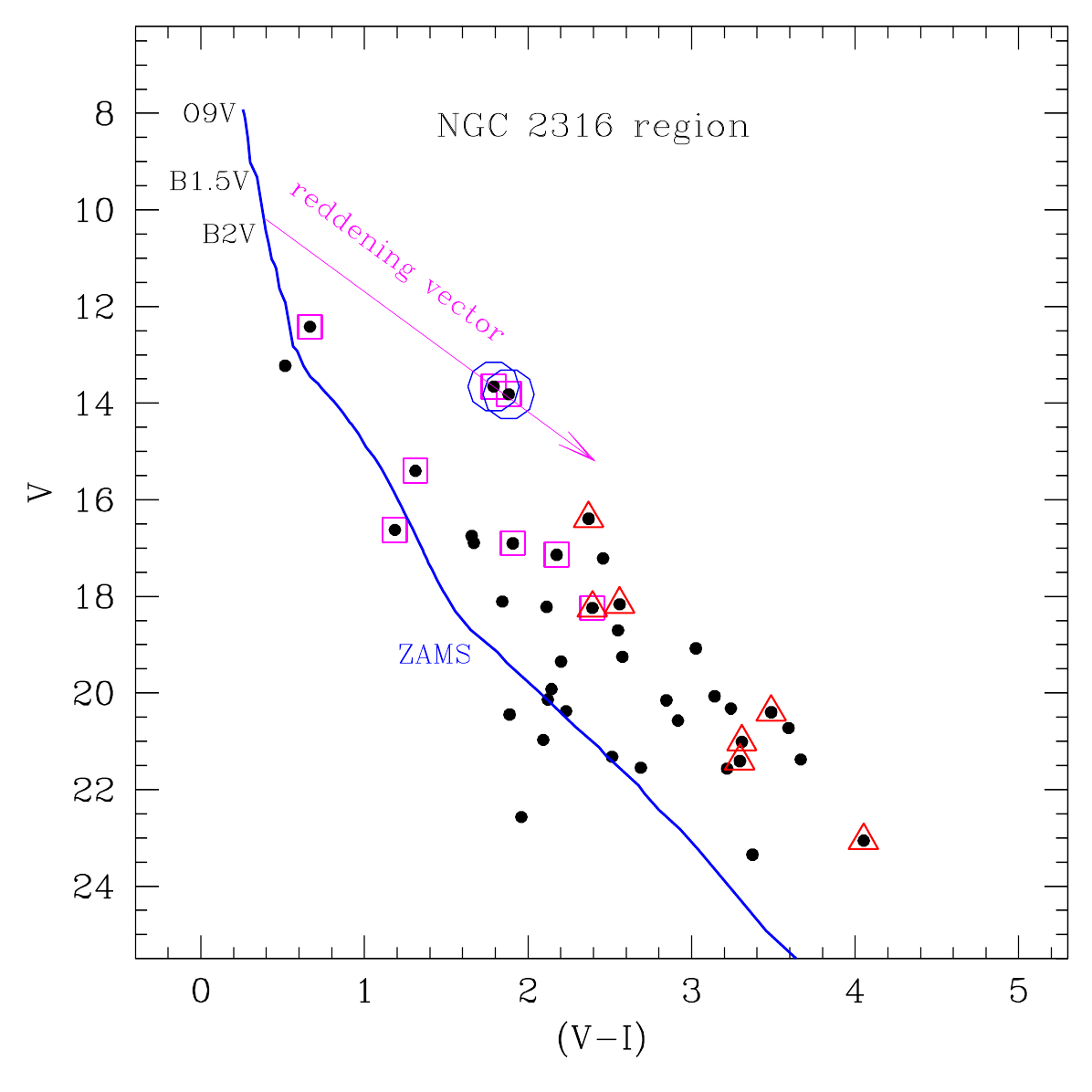}
\caption{\label{tcd} Left panel: $(U-B)$ vs. $(B-V)$ TCD for the sources in the NGC 2316 cluster region (stars within the \emph{convex hull}, black dots). The solid blue and dotted red curves represent the intrinsic ZAMS for $Z=0.02$ given by 
    \citet{2013ApJS..208....9P} which shifted along the reddening vector for $E(B-V)$ = 0.5 mag and  $E(B-V)$ = 1.3 mag, respectively. Right panel: $V$ vs. $(V-I_c)$ CMD for the stars inside the NGC 2316 cluster region (black dots). The blue solid curve represents the ZAMS corrected for the distance 1.3 kpc and reddening $E(B-V) = 0.5$ mag. 
    The candidate massive stars, young stellar objects (Class II), and  \emph{Gaia} DR3 members are shown in both panels by blue circles, red triangles, and square boxes, respectively.}
\end{figure*}

\begin{figure}
\centering
\includegraphics[width=0.45\textwidth]{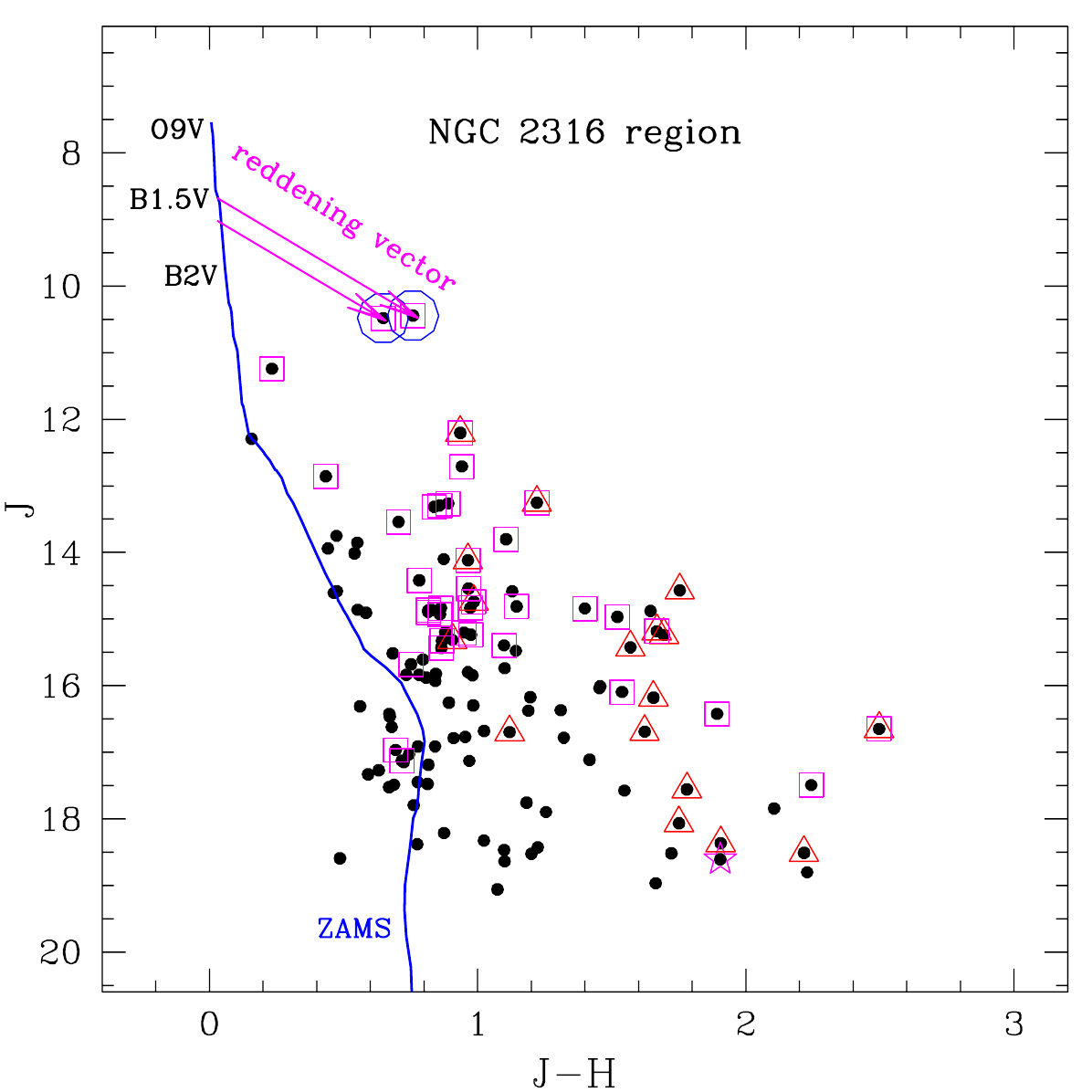}
\caption{\label{ntcd} $J$ vs. $(J-H)$ TCD for the sources in the NGC 2316 cluster region (stars within the \emph{convex hull}). The blue curve represents the intrinsic ZAMS for $Z=0.02$ given by 
\citet{2013ApJS..208....9P} scaled for $E(B-V)$ = 0.5 mag and distance of 1.3 kpc. Symbols are the same as in Figure \ref{tcd}. Asterisks are Class I YSOs.}
\end{figure}

\section{Result and Analysis}\label{sec3}

\subsection{Physical properties of NGC 2316 cluster} \label{group}

To understand the structure of the NGC 2316 cluster, we generated the stellar surface density map of the region by nearest neighbor (NN) method \citep{2005ApJ...632..397G}. This was done in a grid having a size of 10$\arcsec$ and for the twentieth nearest star, using the NIR catalog \citep[for details, please refer, ][]{2023ApJ...953..145V}.
We overlaid the estimated isodensity contours (red curves) in Figure \ref{3panel}. The lowest level for isodensity contours is 24 stars arcmin$^{-2}$ with a step size of 4 stars arcmin$^{-2}$. An apparent clustering of stars having a slightly elliptical geometry is visible inside the ring of dust and gas, peaking at $\alpha_{J2000}=06^h59^m41^s$.0 and $\delta_{J2000}=-07^{\circ}46\arcmin36\arcsec$.
Due to this elliptical morphology, we reconsider the area of this cluster through its \emph{convex hull}\footnote{\emph{Convex hull} is a polygon enclosing all points in a grouping with internal angles between two contiguous sides of less than 180$^{\circ}$.} instead of its circular area since it generally overestimates the area of an elongated cluster \citep{2006A&A...449..151S,2020MNRAS.498.2309S}.  The \emph{convex hull} for the NGC 2316 cluster is generated from the position lowest/outmost isodensity contour (cf. left panel of Figure \ref{3panel}) and is shown in the right panels of Figure \ref{3panel}.
The area of the cluster ($A_{cluster}$ = 3.334 arcmin$^2$) is estimated from the area of the \emph{convex hull} ($A_{hull}$) normalized by a geometric factor \citep[cf. for details,][]{2020MNRAS.498.2309S}, given as:

\begin{equation}
    \begin{split}
        A_{cluster}=\frac{A_{hull}}{1-\frac{n_{hull}}{n_{total}}}
    \end{split}
\end{equation}

where, $n_{hull}$ is the total number of vertices on the hull and $n_{total}$ is the total number of objects inside the hull. The aspect ratio $\frac{R^2_{circ}}{R^2_{cluster}}$
of this cluster is 1.6, where $R_{cluster}$ (= 1$\arcmin$.03 $\sim$ 0.4 pc) is the radius of the circle whose area is equal to $A_{cluster}$ and $R_{circ}$ (= 1$\arcmin$.3 $\sim$ 0.5 pc) is half of the farthest distance between two hull objects. 
\citet{2004A&A...413L...1T} overestimated the size of this cluster (0.63 pc) from its radial profile, which does not consider the cluster's elongation.

As the NGC 2316 cluster seems to be embedded in a nebulosity of gas/dust, it is essential to estimate the extinction of this cluster to determine the other physical parameters, such as distance, age, etc. We used $(U-B)$ vs. $(B-V)$ two-color diagram (TCD) for this purpose, which is shown in the left panel of Figure \ref{tcd}. We have shown the distribution of stars in the NGC 2316 cluster, i.e., those lying within the \emph{convex hull} (as shown in Figure \ref{3panel}) as black dots in the TCD.
We also plot the intrinsic zero-age main sequence (ZAMS; blue curve), which we took from \citet{2013ApJS..208....9P}.
We have also over-plotted the distribution of probable member stars (identified based on $Gaia$\footnote{https://gea.esac.esa.int/archive/} Proper Motion (PM) data, cf. Appendix \ref{memberg}) by square symbols.  The PMS stars showing excess infrared (IR) emission \citep[for details on methodology, please refer][]{2023ApJ...953..145V} are also shown by triangles. 
The minimum reddening towards the cluster can be determined by visually fitting the ZAMS to the bluer end of spectral type A distribution stars or earlier. Various aspects, such as the distribution of binary stars, pre-main-sequence (PMS) stars, metallicity, and photometric errors, have also led us to such considerations (refer \citealt{2006AJ....132.1669S,1994ApJS...90...31P} for further details). We shifted the ZAMS along the reddening vector having a slope of $E(U-B)/E(B-V)$ = 0.72 (for normal Galactic reddening law `$R_V$ = 3.1', \citealt{1979A&AS...38..197M,1989AJ.....98..611G}) to the distribution of stars (blue curve in the left panel of Figure \ref{tcd}), hence estimated minimum reddening towards the cluster as $E(B-V)_{min}$ = 0.50 $\pm$ 0.05 mag (corresponding to $A_V=1.55\pm0.20$ mag, for normal Galactic reddening law i.e., `$R_V$ = 3.1', \citealt{1979A&AS...38..197M,1989AJ.....98..611G}). The approximate error in the reddening measurement is calculated by the steps given in \citet{1994ApJS...90...31P}. The probable massive stars (blue circles, please refer to their position in the optical color-magnitude diagram; CMD) seem to suffer more extinction ($E(B-V)$ = 1.3 mag or $A_V$ = 4.0 mag, red dotted curve) above the minimum extinction value derived for the cluster. 

The distance of the NGC 2316 cluster was estimated as 1.1 kpc by \citet{1994LNP...439..175F}.  To validate the distance of this cluster, we used the mean of the distances reported by \citet{2021AJ....161..147B} of the 18 cluster members (refer Appendix \ref{memberg} for detailed investigation on cluster membership), having a membership probability $\geq 80 \%$ and the parallax values with error $\leq 0.4$ mas. We found the distance of this cluster as 1.3 $\pm$ 0.3 kpc. We also used the $V$ vs. $(V-I_c)$ CMD from our deep optical observations (right panel of Figure \ref{tcd}) of the stars within the cluster to confirm this distance and determine the age of cluster (\citealt{2020ApJ...891...81P, 2020ApJ...896...29K}). The probable massive star, probable members of the NGC 2316 cluster, and the identified young stellar objects (YSOs) are also marked by blue circles, magenta squares, and red triangles, respectively, in the right panel of Figure \ref{tcd}. The intrinsic ZAMS (blue curve) taken from \citet{2013ApJS..208....9P}, is also plotted for an extinction $E(B-V)_{min}$ = $0.5$ mag and a distance of 1.3 kpc. The ZAMS appears very well fitted to the blue envelope of the stars' stars and hence confirms the distance and extinction estimations (for detailed CMD isochrones fitting, see \citealt{1994ApJS...90...31P}).  The location of the probable massive stars in the CMD suggests their spectral type to be around B1.5V-B2V. Since these are the brightest member stars of the NGC 2316 cluster, the upper age limit of this cluster is restricted to $\sim$12 Myr \citep{2004fost.book.....S}.
Some of the member stars fall in the PMS stage in the CMD, and some of the stars show excess infrared (IR) emission (i.e., stars with discs around them, age $<$ 3 Myr \citep{Evans_2009}); we conclude that NGC 2316 is still forming young stars despite the fact that the most massive stars were born $\sim$12 Myr ago.
For further confirmation, we plotted the NIR CMD for the stars in the NGC 2316 cluster region in Figure \ref{ntcd}. 
As expected, the NIR observations detected many more stars than the optical one; thus, we can see more stars deeply embedded in the nebulosity of the region. Most of the embedded stars are very young, with disks around them.

\begin{figure*}
\centering
\includegraphics[width=0.45\textwidth]{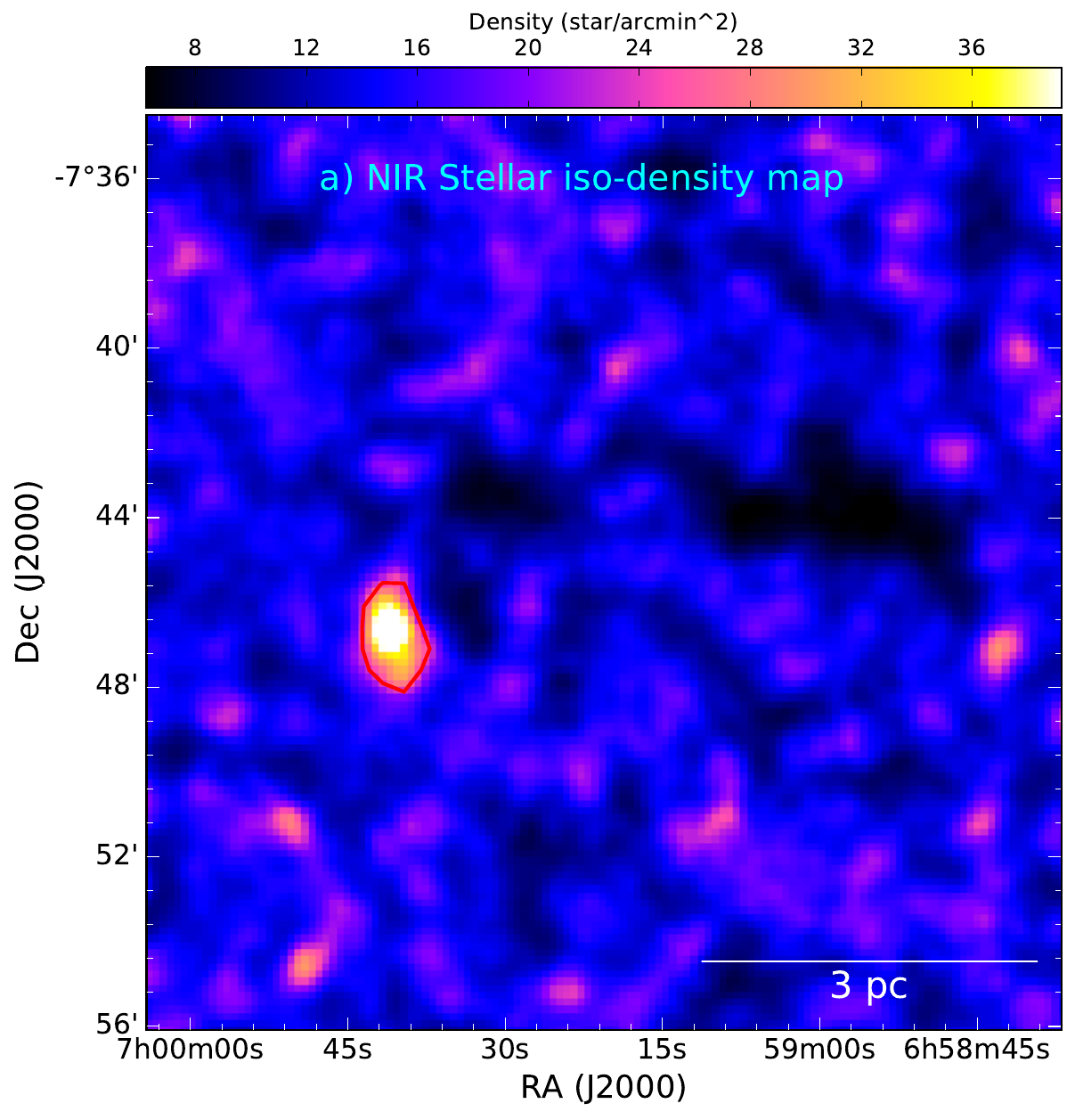}
\includegraphics[width=0.45\textwidth]{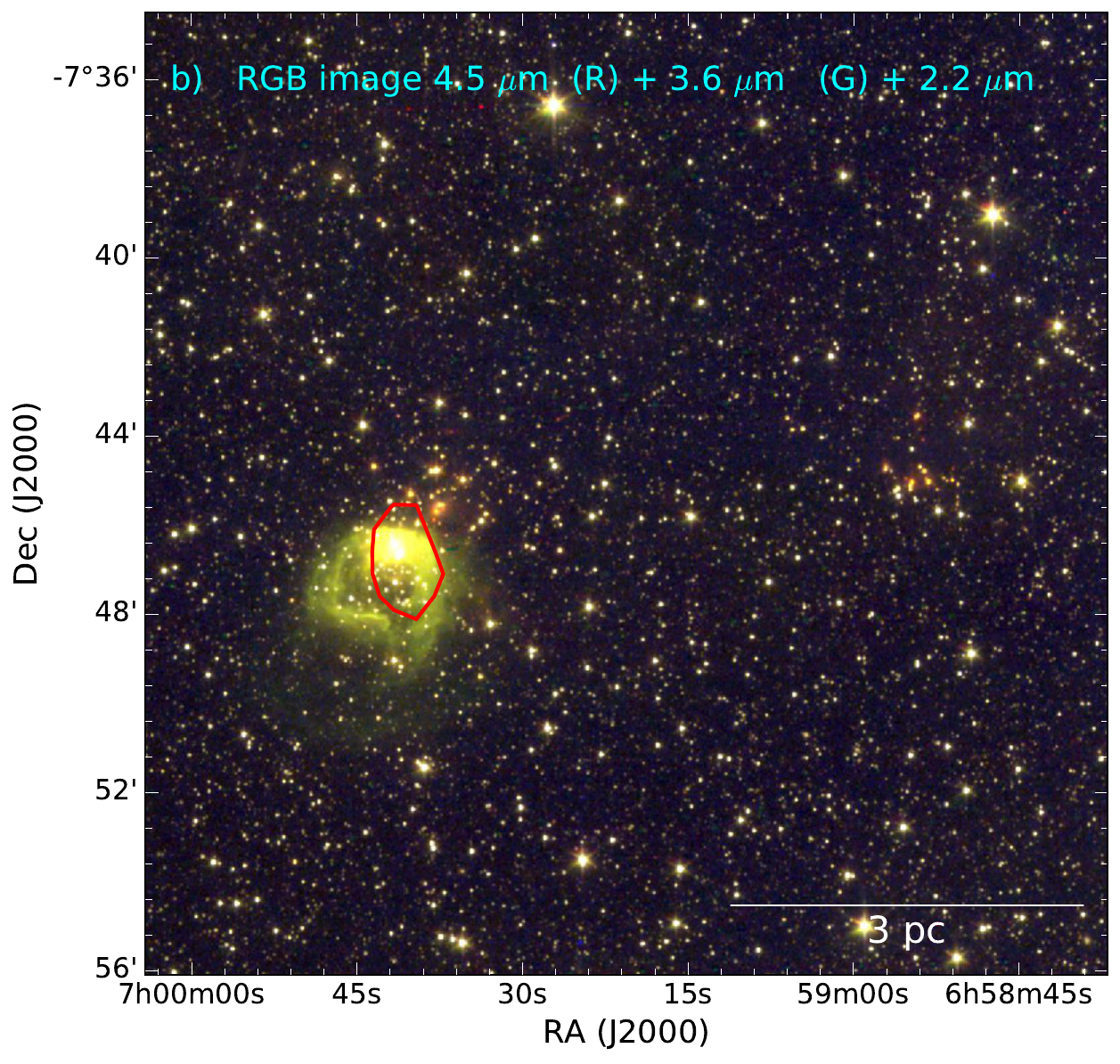}
\includegraphics[width=0.45\textwidth]{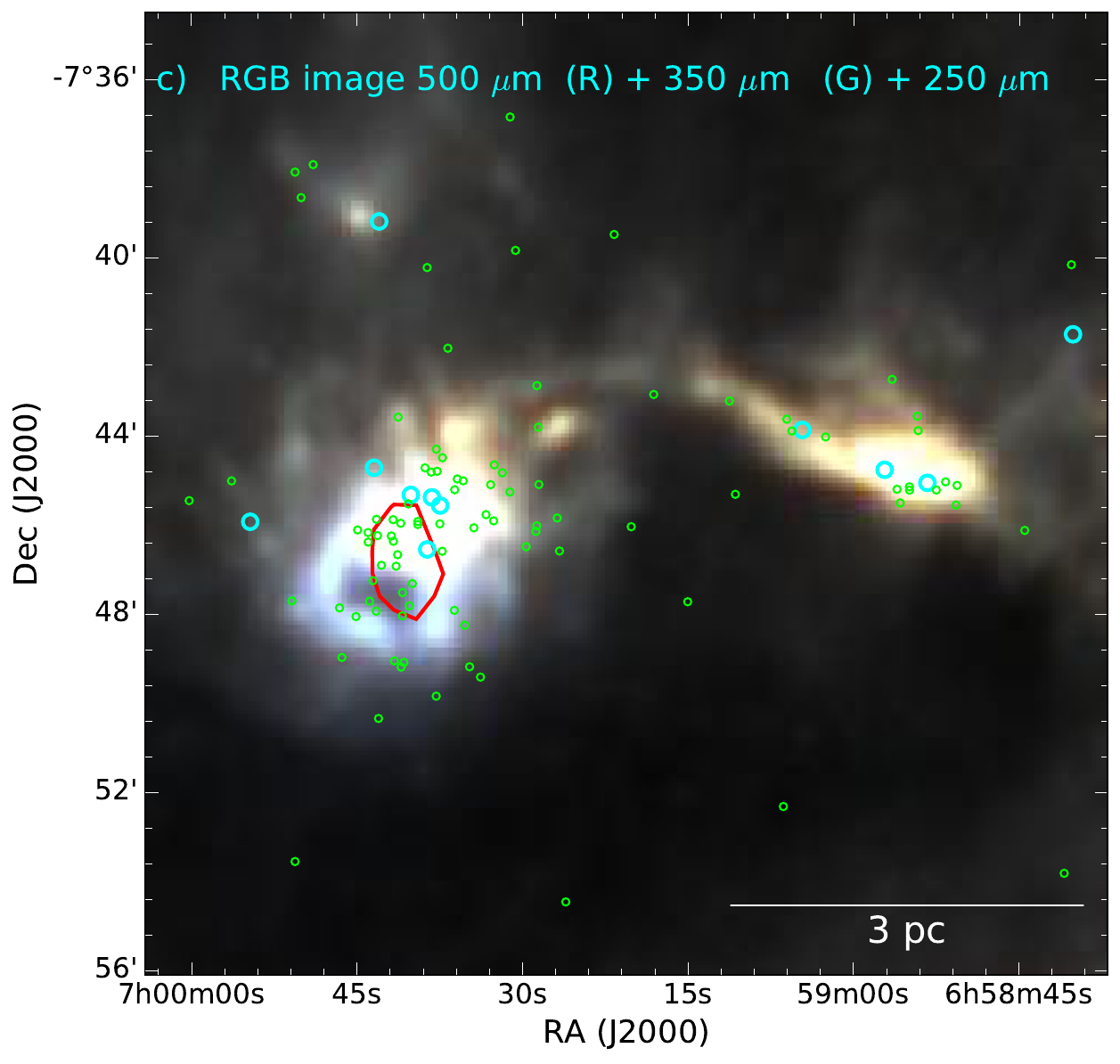}
\includegraphics[width=0.45\textwidth]{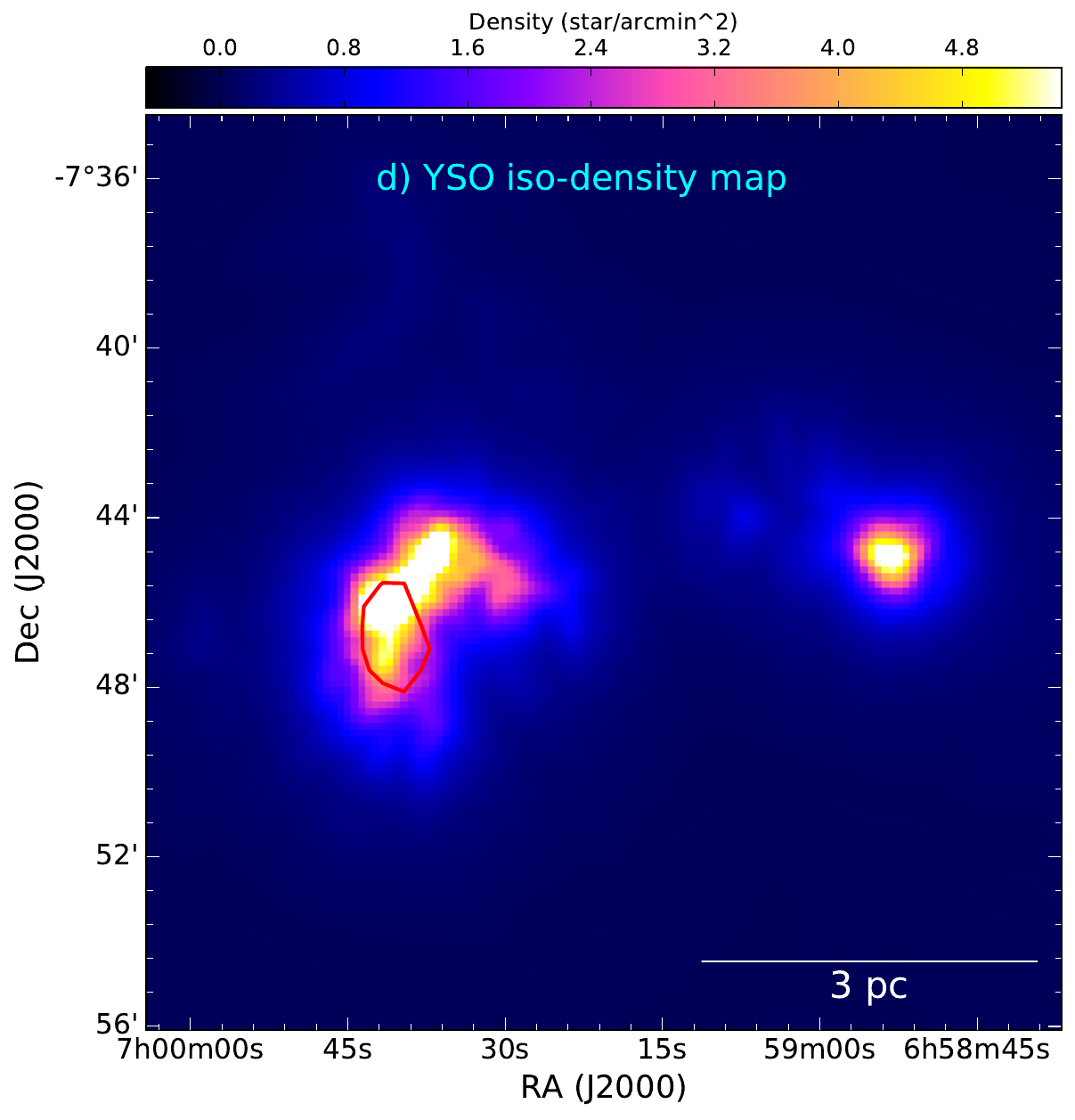}
\caption{\label{ir} Panel (a): Stellar density map derived from the NIR catalog. Panel (b): NIR color-composite image (Red: \emph{Spitzer} 4.5 $\mu$m; Green: \emph{Spitzer} 3.6 $\mu$m; Blue: 2MASS K 2.17 $\mu$m). Panel (c): FIR color-composite image (Red: \emph{Hershel} 500 $\mu$m; Green: \emph{Hershel} 350 $\mu$m; Blue: \emph{Hershel} 250 $\mu$m) overlaid with the locations of Class I (cyan circles asterisk) and Class II (green circles) YSOs.
Panel (d): YSOs density map. The shown FOV is $21^\prime \times21^\prime$ of the NGC 2316 region, and all panels are overlaid with the NGC 2316 cluster region, i.e., \emph{Convex hull} (red polygon).
}
\end{figure*}

\begin{figure*}
\centering
\includegraphics[width=0.47\textwidth]{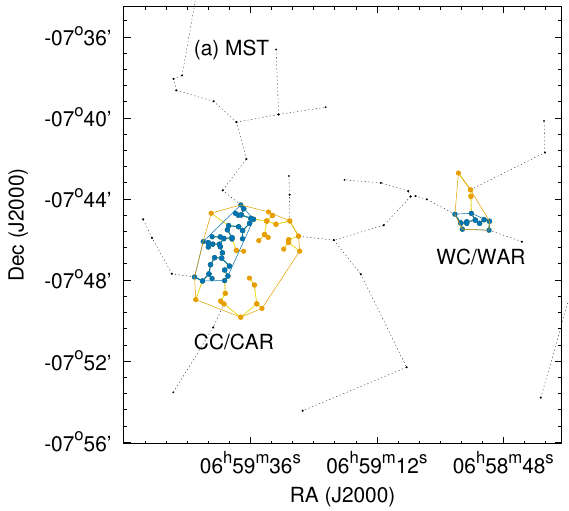}
\includegraphics[width=0.43\textwidth]{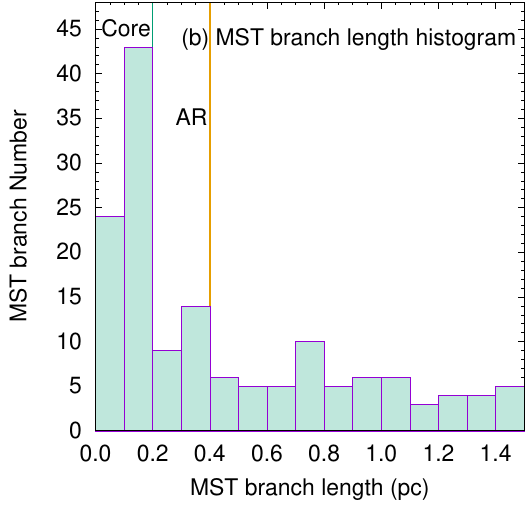}
\includegraphics[width=0.45\textwidth]{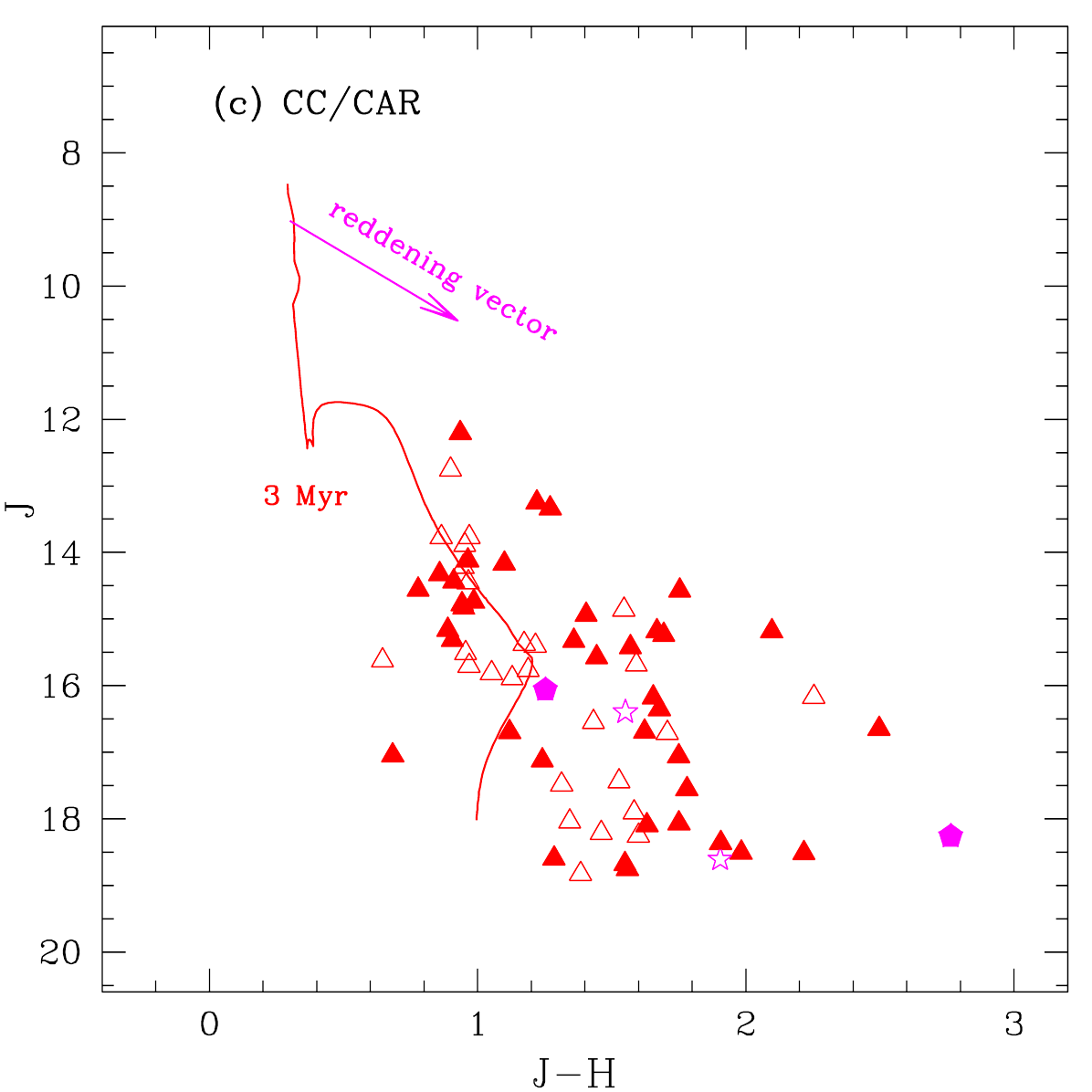}
\includegraphics[width=0.45\textwidth]{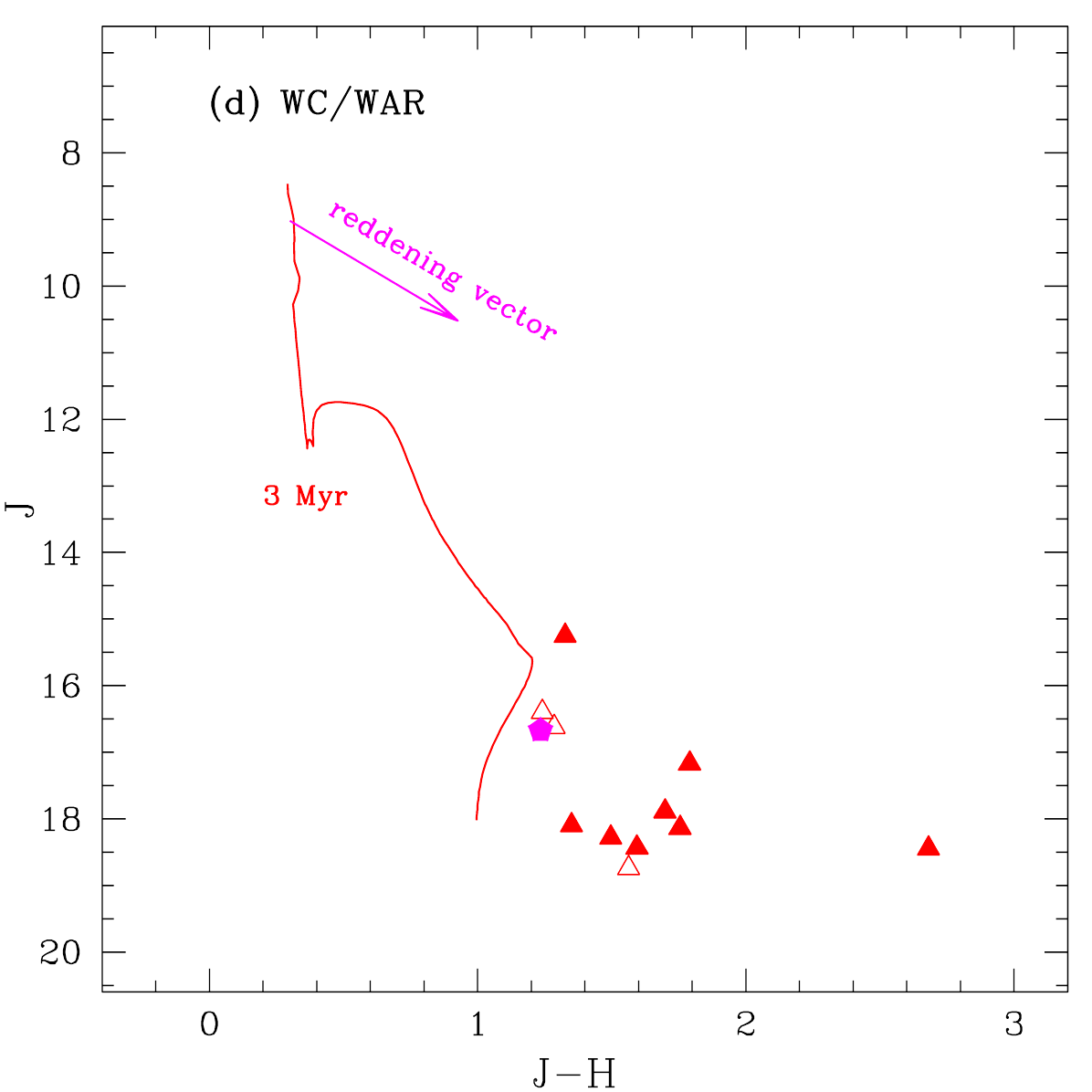}
\caption{\label{mst} Panel (a): MST for the location of YSOs. The isolated cores (blue dots) and ARs (Blue $+$ Yellow dots) are shown by their respective \emph{Convex hulls}. Panel (b): Histogram for the MST branch length.
 Panel (c): $J$ vs. $(J-H)$ CMD for the YSOs located inside the CC (filled triangles (Class II)/filled asterisks (Class I)) and CAR (triangles (Class II)/asterisks (Class II)). 
 Panel (d): Same as panel (c) but for WC and WAR. The red solid curve in panels (c) and (d) represents the isochrone of 3 Myr and $Z=0.02$ given by \citet{2019MNRAS.485.5666P}, which is shifted along the reddening vector for $A_V$ = 4 mag and the distance of 1.3 kpc.
}
\end{figure*}

\begin{figure}
\centering
\includegraphics[width=0.45\textwidth]{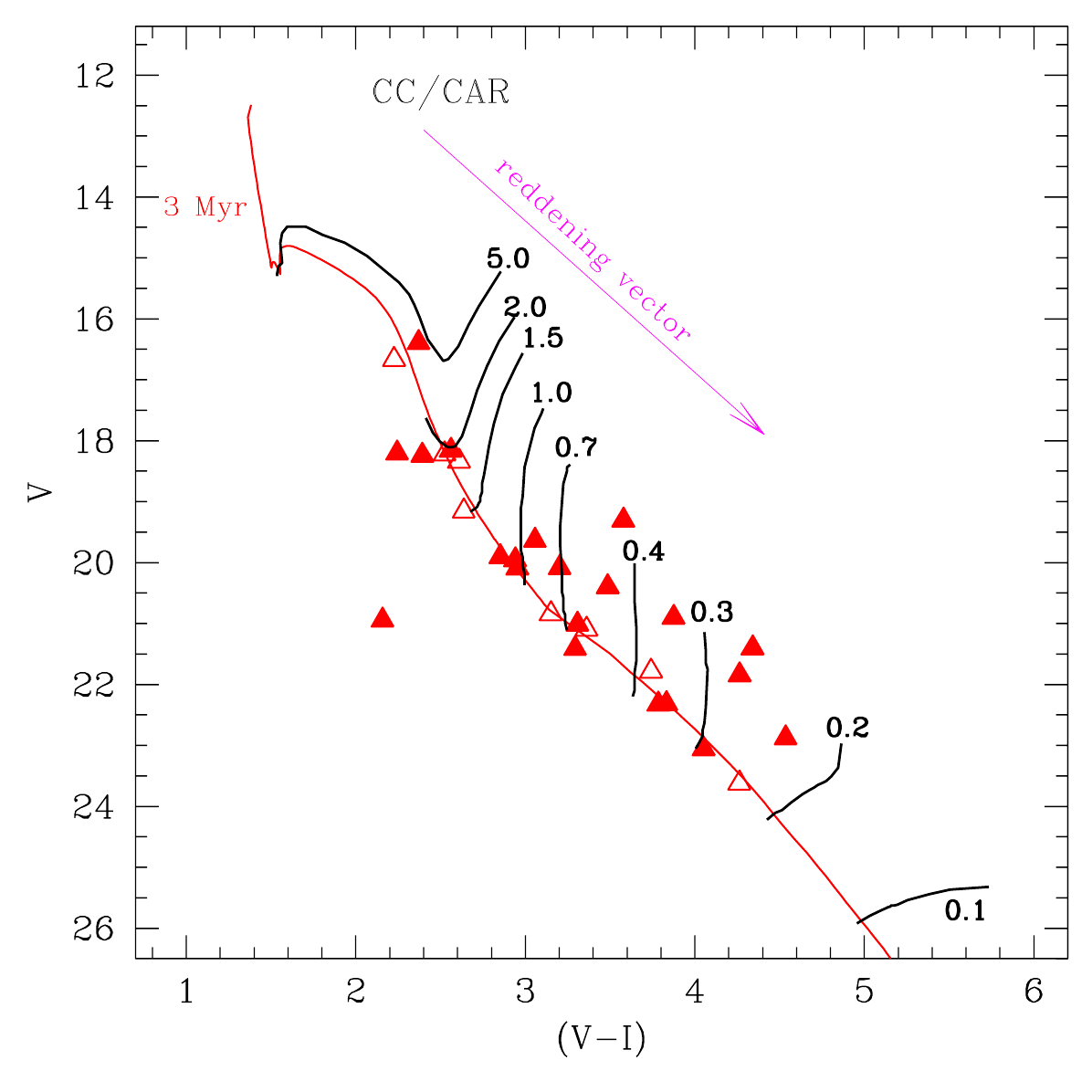}
\caption{\label{tcdyso}  $V$ vs. $(V-I_c)$ CMD for the YSOs (Class II) located inside the  CC (filled triangles) and CAR (triangles).  The red solid curve represents the isochrone of 3 Myr and $Z=0.02$ given by  \citet{2019MNRAS.485.5666P}, which is shifted along the reddening vector for $A_V$ = 4 mag and the distance of 1.3 kpc. Black lines are the evolutionary tracks for different mass stars.
}
\end{figure}

\begin{table*}
\centering
\scriptsize
\caption{\label{Tp1} Properties of the identified cores and ARs.
The center coordinates are given in columns 2, 3. The hull radius, circle radius, and aspect ratio are given in columns 4, 5, and 6, respectively. Columns 7 and 8 represent the mean and peak stellar density obtained using the isodensity contours from the NIR catalog. Columns 9, 10, and 11 are the median MST branch length, mean, and peak extinction values.  Column 12 represents the cloud mass in the \emph{convex hull} derived using the extinction map. Columns 13, 14, and 15 represent the number of YSOs, their $Q$ parameters, and the fraction of Class I YSOs over total YSOs, respectively. Columns 16 and 17 represent YSOs' mean age and mean mass derived from the spectral energy distribution (SED) analysis. The last column is each region's star formation efficiency (SFE).
}
\begin{tabular}{@{}l@{}cc|cc@{}ccccccc|ccc@{}c@{}c@{}|c@{}}
\hline
Region& $\alpha_{(2000)}$&$\delta_{(2000)}$& $R_{\rm hull}$& $R_{\rm cir}$& Aspect & $\sigma_{\rm mean}$&$\sigma_{\rm peak}$ & MST$^a$  &  $A_{V_{mean}}^b$ & $A_{V_{peak}}$ & Mass &N$^c$& $Q$&Frac$^d$& Age & Mass& SFE \\
 & {\rm $(^h:^m:^s)$} & {\rm $(^o:^\prime:^{\prime\prime)} $} &  (pc)& (pc)& Ratio & (pc$^{-2}$)& (pc$^{-2}$)  &  (pc) &     (mag)          & (mag) &  ($M_\odot$) &   &   &(\%)& (Myr) & ($M_\odot$)& (\%)\\
\hline
CC &   6:59:40.8  & -7:46:12 &    0.57 & 0.79 & 1.9 & 38.7 & 110.8&0.12 &7.3&10.4 & 88 &  36 & 0.66& 5 & 1.8 & 1.5 &38\\
CAR &  6:59:38.2  & -7:46:29 &    1.06 & 1.13 & 1.1 & 20.8 & 110.8&0.14 &6.1&10.4 &  258&  61& 0.78& 6& 1.8 & 1.2 &22\\
&&&&&&&&&&&&&&&&&\\
WC  &  6:58:53.8  & -7:45:08 &    0.30 & 0.34 & 1.2 & 36.4 & 68.3&0.12  &5.2&5.8 & 13 &  11& 0.79& 13 & 1.5 & 0.8&41\\
WAR  & 6:58:54.0  & -7:44:45 &  0.45 & 0.61 & 1.8 & 22.9 & 68.3&0.12  &5.5&7.3 & 30 &  14& 0.75 & 8 &1.3 &0.7&24\\
\hline
\end{tabular}

a: median branch length;
b: from extinction map from NIR data;
c: Number of YSOs enclosed in the group;
d: Class I/(Class I + Class II)

\end{table*}

\subsection{Physical environment around NGC 2316 cluster}\label{sec:physical_environment}

The recent archival wide-field IR, MIR, and far-infrared (FIR) surveys can provide a clear picture of the distribution of young stars, dust, gas, etc., which will help us to explore the star formation scenario more \citep{2010A&A...523A...6D, 2017ApJ...834...22D}. The panel (a) of Figure \ref{ir} depicts the stellar density maps generated from our NIR catalog over-plotted with the \emph{convex hull} for the NGC 2316 cluster. While we can see the overdensity of stars in the NGC 2316 cluster region, there is a dark lane (absence of stellar density) starting near the cluster region, extending towards the western direction. In the panel (b) of Figure \ref{ir}, we show the color-composite image generated using the \emph{Spitzer} 4.5 $\mu$m (red), 3.6 $\mu$m (green), and 2MASS 2.2 $\mu$m (blue) band images. The \emph{Spitzer}/2MASS images manifest warm dust and/or photospheric stellar emissions. We can see the distribution of stars throughout the region and warm gas/dust distribution towards the NGC 2316 cluster region. The warm gas/dust shape represents a bubble near the NGC 2316 cluster. The bubble's center and the cluster seem to be slightly off-centered. There is also a distribution of red point sources at the top of the cluster and towards the western end of the dark lane of stellar density.
In the panel (c) of Figure \ref{ir}, we show the color-composite image generated using the \emph{Hershel} 500 $\mu$m (red), 350 $\mu$m (green), and 250 $\mu$m (blue) band images.  
The \emph{Herschel} FIR images represent the region's colder gas/dust distribution. Thus, FIR images manifest a ring or a shell-like structure near the NGC 2316 cluster, along which a long lane starts from the cluster to the western direction. This lane more or less represents the dark region of the density map, which is evident because of the high extinction of the material. 
The FIR emission peaks at the western end of the dark lane, where we have seen the distribution of red sources. As this region shows many signatures of star-forming activities, we can expect young populations of stars to be embedded in this region. Since these young populations (or YSOs) are usually associated with the circumstellar disc, we identified/classified them based on their excess IR-emission \citep[for details, please refer][]{2023ApJ...953..145V}. We found both Class I and Class II YSOs in our selected FOV of this region and their locations are shown in panel (c) of Figure \ref{ir}. The locations of YSOs mostly follow the FIR emission. The region has a larger spread of colder gas/dust, followed by the distribution of YSOs. 
Here, it is worthwhile to note that younger Class I YSOs are located either just outside the boundary of the NGC 2316 cluster region or towards the western end of the dark lane, having a peak in the FIR emission. We have also generated the density maps of the YSOs, similar to the density maps generated from the NIR catalog in Section \ref{group}. Panel (d)  of Figure \ref{ir} shows the YSOs density maps overlaid with the \emph{convex hull} of the cluster region. From this image, we can conclude that the YSOs distribution is more extended than the distribution of cluster stars. The YSOs show two density peaks, one towards the cluster region and another towards the western end of the gas/dust lane.

\begin{figure*}
\centering
\includegraphics[width=0.47\textwidth]{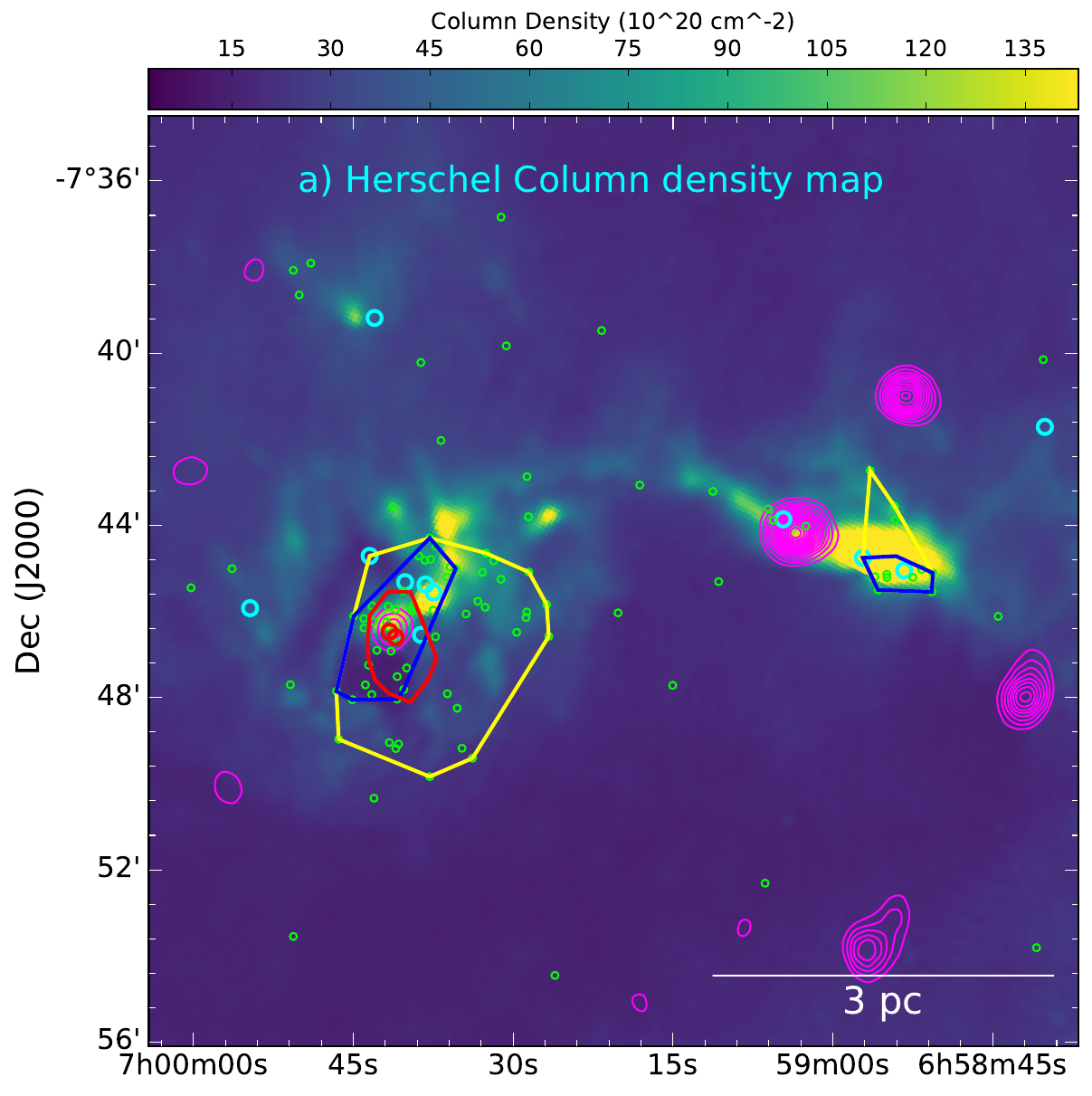}
\includegraphics[width=0.45\textwidth]{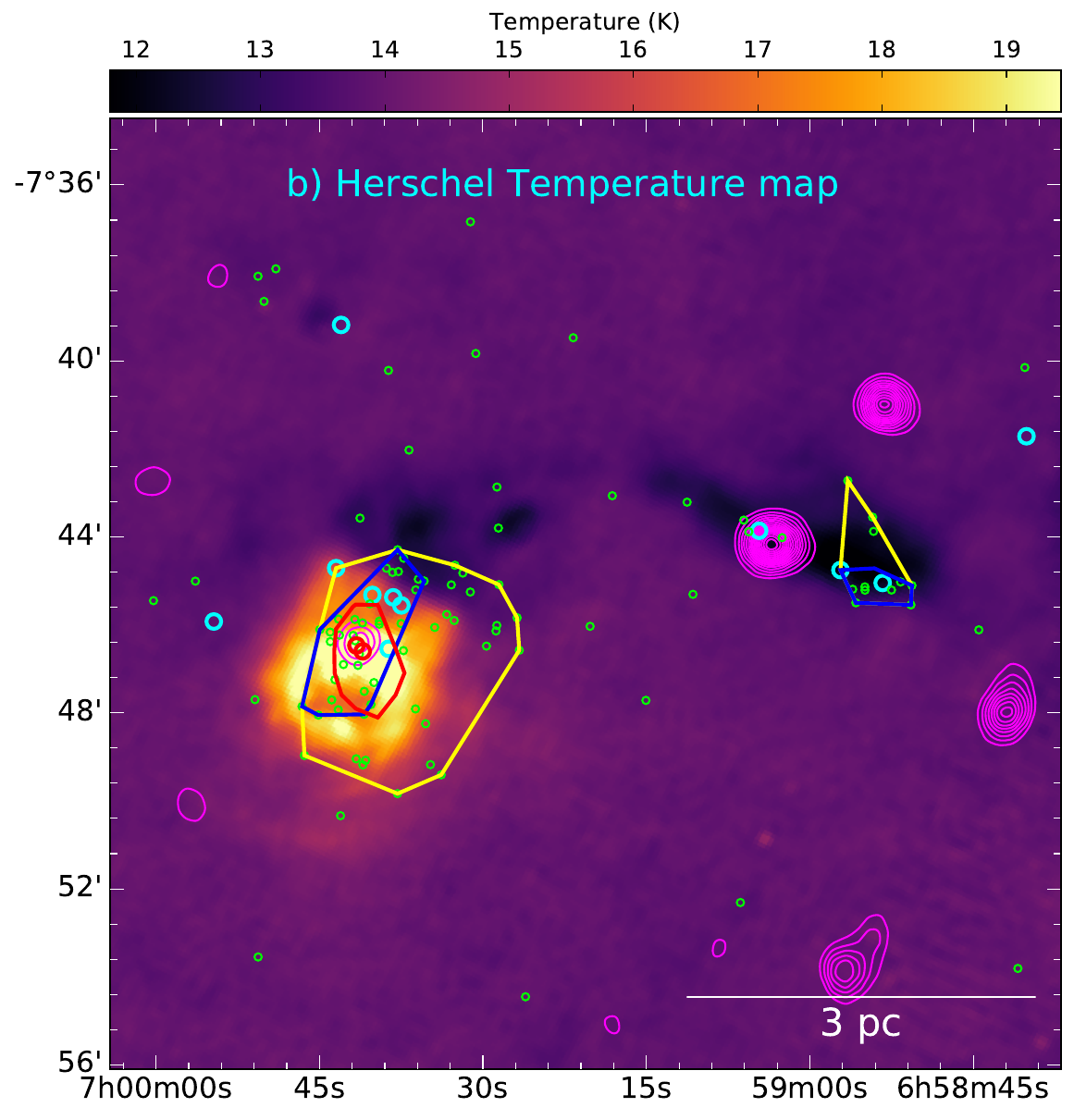}
\includegraphics[width=0.47\textwidth]{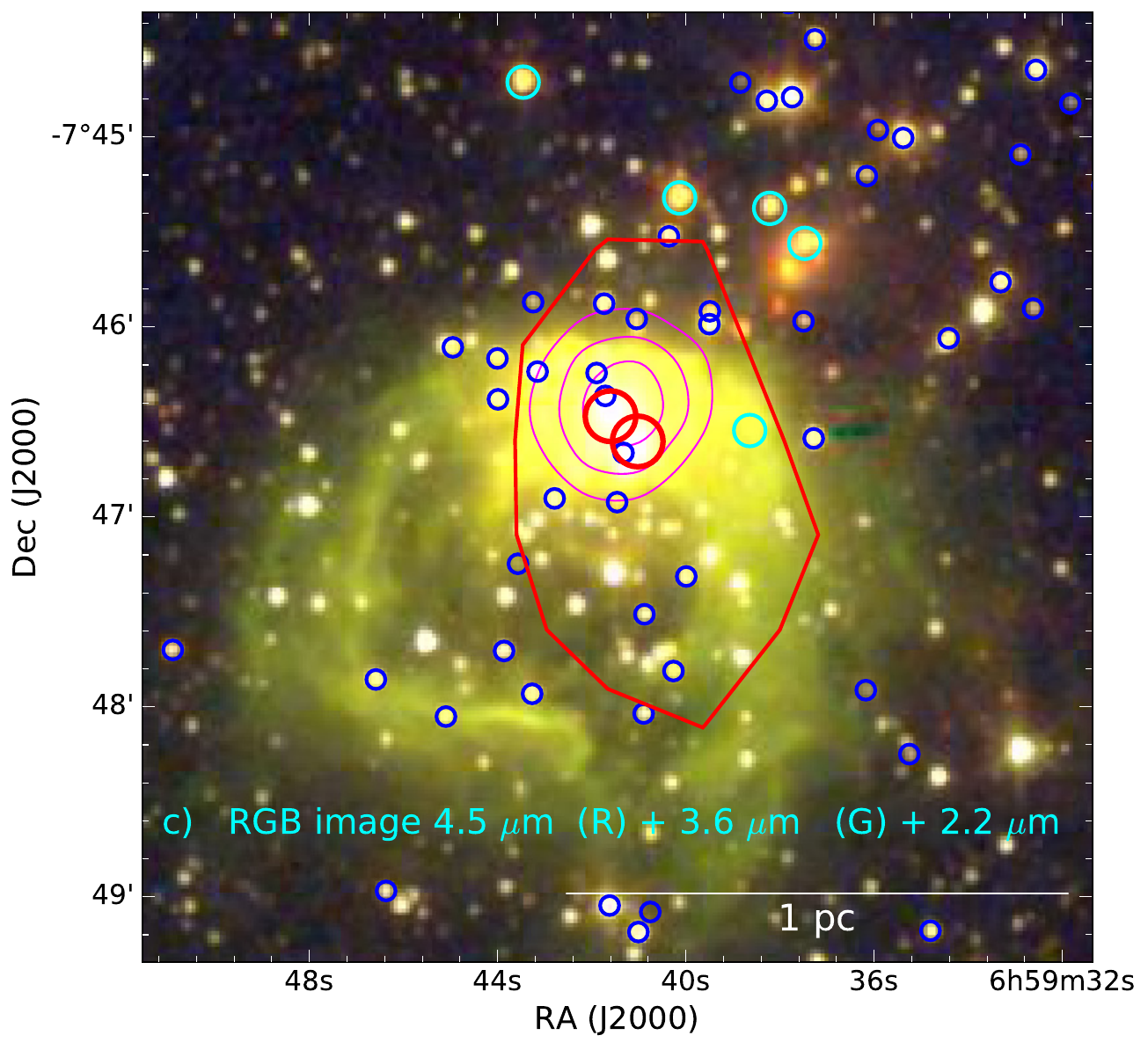}
\includegraphics[width=0.45\textwidth]{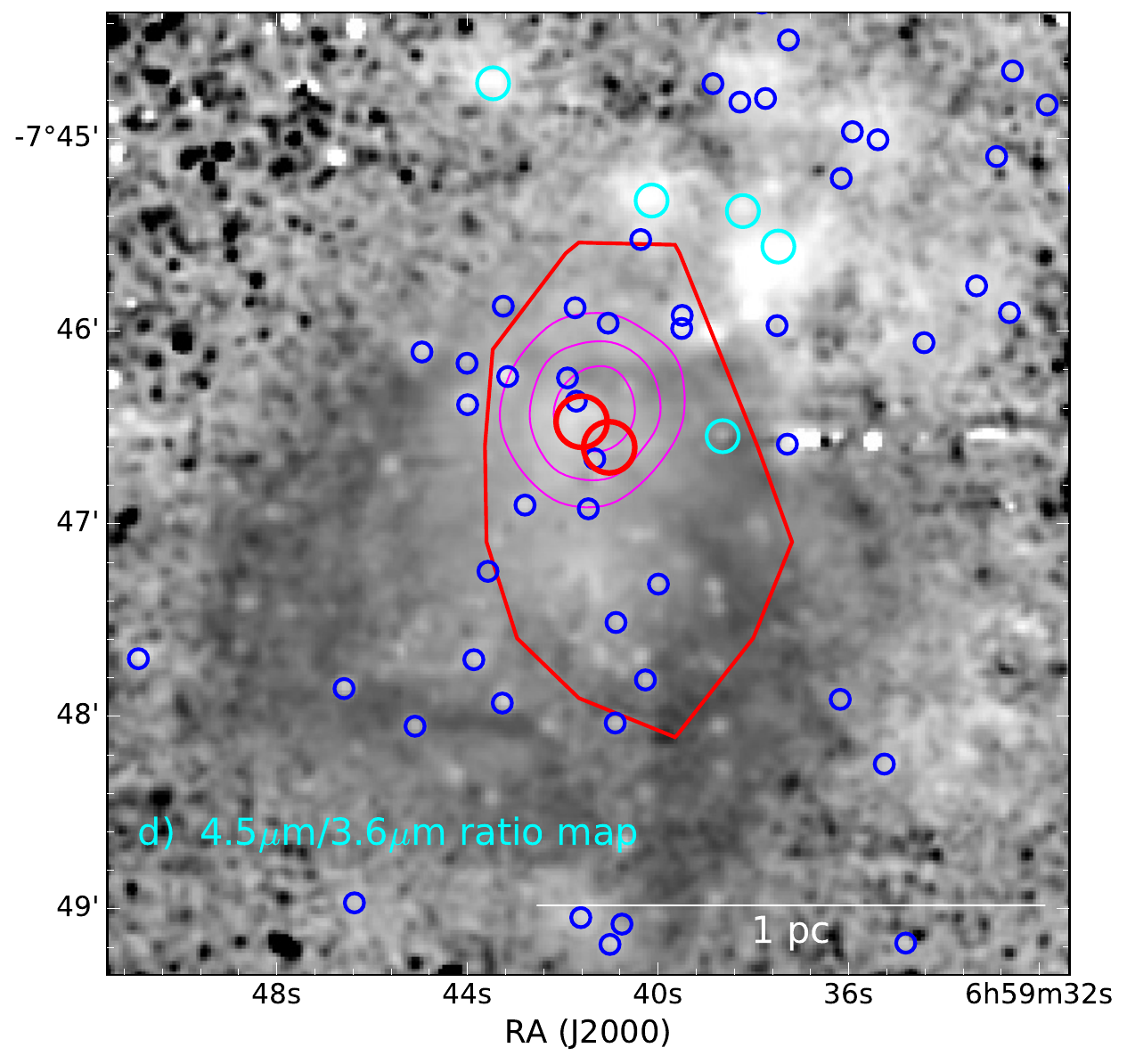}
 \caption{ \label{massive}  Panel (a):  \emph{Herschel} column density map. Panel (b): \emph{Herschel} temperature map. Panel (c): Color-composite image (Red: \emph{Spitzer} 4.5 $\mu$m; Green: \emph{Spitzer} 3.6 $\mu$m; Blue: \emph{UKIDSS} 2.2 $\mu$m). Panel (d): \emph{Spitzer} ratio map of 4.5 $\mu$m/3.6 $\mu$m emission, smoothed using Gaussian function. The ratio map overlaps the NVSS 1.4 GHz radio continuum contours (magenta). The lowest level to generate these contours is 1.5 mJy\,beam$^{-1}$ with a step size of 1 mJy\,beam$^{-1}$.  
All the panels show the location of the identified massive stars (red circles) and the cluster region (red polygon). Panels (a), (b), and (c) also show the location of Class I (cyan circles) and Class II (green/blue circles) YSOs. The \emph{convex hull} for identified  YSO cores and ARs are also shown in Panel (a) and (b). Panel (c) and (d) are zoomed-in images for the NGC 2316 cluster region.
}
\end{figure*}

\subsection{YSO population in the region}\label{sec:extraction_of_cores}

To isolate the YSOs sharing similar star formation history, we applied an empirical technique \emph{`Minimal Spanning Tree'} (MST; \citealt{2009ApJS..184...18G}).
It is one of the best techniques as it isolates the groupings without any bias or smoothing and preserves underlying geometry \citep{2004MNRAS.348..589C,2006A&A...449..151S,2009MNRAS.392..868B,2009ApJS..184...18G}. The methodology for the extraction of MST is discussed in our previous publications, i.e., \citet{2016AJ....151..126S,2017MNRAS.467.2943S,2020ApJ...891...81P}, and is plotted in the panel (a) of the Figure \ref{mst} for the identified YSOs. In the panel (b) of Figure \ref{mst}, we show the distribution of MST branch length. Clearly, it shows a peak around 0.2 pc. We have shown those YSOs in the left panel with blue dots. Also, the branch length distribution shows a second peak at 0.4 pc. We have shown those YSOs by yellow dots in the left panel. We call the first distribution the core region, whereas the second is the active region (AR). We have also generated their respective \emph{convex hulls} as shown in panel (a) of Figure \ref{mst}. Thus, we have successfully isolated the two separate cores and ARs in the NGC 2316 complex from a diffuse distribution of YSOs. We have named these as central core (CC), central AR (CAR), western core (WC), and western AR (WAR), as shown in panel (a) of Figure \ref{mst}. 
We have derived the physical parameter of these cores/ARs as listed in Table \ref{Tp1} using the methods explained in \citet{2016AJ....151..126S,2017MNRAS.467.2943S}.

We show the distribution of YSOs associated with the CC/CAR and WC/WAR in the NIR CMDs as shown in panel (c) and panel (d) of Figure \ref{mst}, respectively. As NIR data can detect deeply embedded sources, we can see a much richer population of YSOs in these CMDs. We have also overplotted the 3 Myr isochrone (approximate upper age limit of disc-bearing YSOs) in the CMDs. The CC YSOs have a massive population in comparison to the WC YSOs. Also, the WC/WAR YSOs seem much more deeply embedded in the nebulosity. This is also why we could not find any optical counterparts of YSOs in the WC/WAR. Thus, in Figure \ref{tcdyso}, we show the optical CMD of the YSOs in CC/CAR along with the evolutionary sequence for various mass stars.
The CC/CAR YSOs mostly follow the 3 Myr isochrone with a mass between $\sim$ 0.2 to 5 M$_\odot$. 

\begin{figure*}[!t]
\centering
\includegraphics[width=\textwidth]{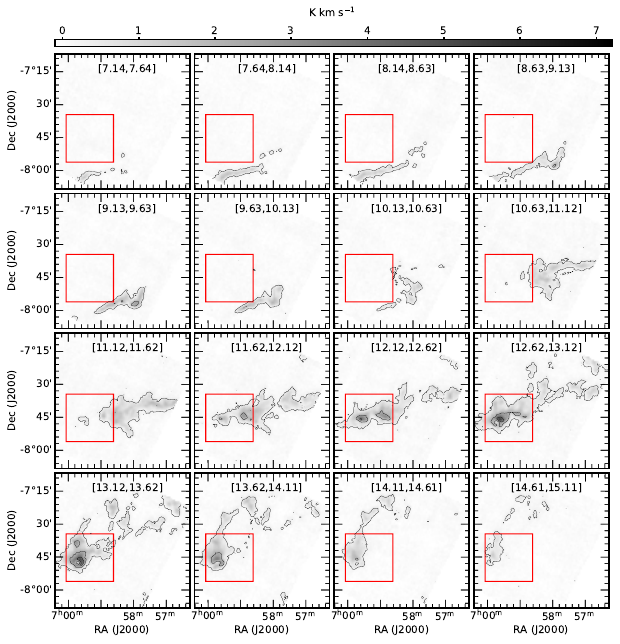}
\caption{
\tco\, Channel Map of the larger region. Two nearly parallel SE-NW elongated 
(filamentary) structures, delineated best in $\sim$[8.63, 9.13]\kms\, and $\sim$[12.12, 12.62]\kms\, channels respectively, can be seen. The redshifted (filamentary) cloud shows many filaments joining it. The red box is our targeted region as shown in the panel (a) of Figure \ref{3panel}.}
\label{fig_13CO_ChannelMap}
\end{figure*}

\begin{figure*}[t]
\centering
\includegraphics[width=\textwidth]{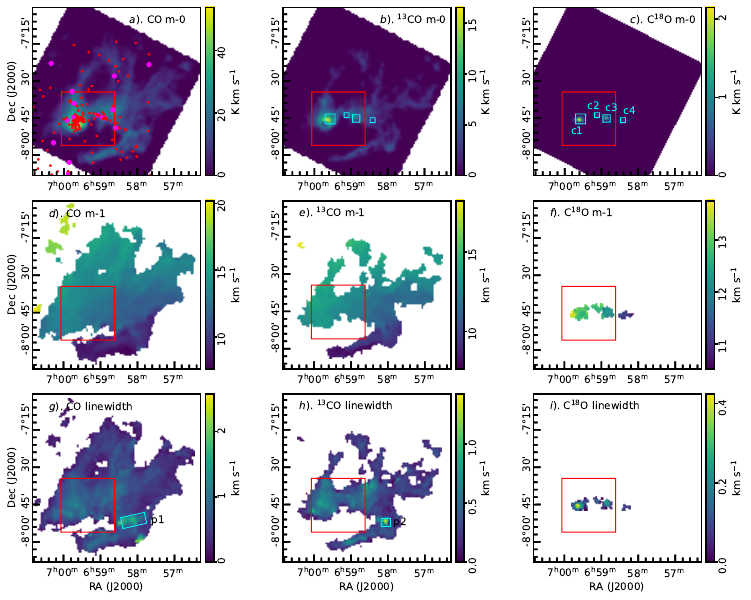}
\caption{
\emph{Column-wise :} \co, \tco, and \cetno\, moment maps (m-0, m-1, and 
linewidth maps in respective rows) constructed using emission regions, which 
display emission above 3$\sigma$ level ($\sigma$ being the rms noise level 
of the respective cubes).
In panels (b) and (c), \cetno\, clumps where spectra were extracted 
(see Figure \ref{fig_CO_spectra}) have been marked by boxes. 
These boxes are labeled in panel (c). 
In panels (g) and (h), the regions of high dispersion have been marked by 
boxes and labeled p1 and p2, respectively. Spectra was extracted at these 
two locations as well. The red box is our targeted region, as shown in panel (a) of Figure \ref{3panel}. Locations of YSOs are also shown on the top-left panel.
}
\label{fig_CO_m012}
\end{figure*}

\subsection{Influence of massive star}\label{sec:pressure}

In panel (a) of Figure \ref{massive}, we present \emph{Herschel} column density map of the region. The location of YSOs/massive stars and the  NVSS 1.4 KHz radio continuum emission contours are also shown in the image. We noticed the radio continuum emission around the probable massive stars inside the bubble of warmed-up gas. There are a few other radio peaks in this region, but they are not associated with warm dust. Thus, we have not considered them for our analysis. The western end of the dust lane having YSOs is not associated with radio emission. The identified YSO cores/ARs and the NGC 2316 cluster region are also shown in this Figure.
The column density map shows the fragmented structure with higher column density near the massive star and a lane extending towards the western direction with a peak at its end. Several dense regions on the map are associated with younger YSOs. The massive stars and the cluster region are associated with the YSO core/AR, i.e., CC/CAR. The western dust/gas peak also hosts YSOs' other core/AR, i.e., WC/WAR.  Most of the Class I YSOs belong to these cores. Here, it is worthwhile to note that the center of the cluster and the location of the massive stars are a bit off from the gas/dust distribution peak. The western YSOs core seems to be associated with a high column density. The YSO distribution is more extended than the cluster but lies mainly within the MIR emission.
The panel (c) of Figure \ref{massive} represents the corresponding temperature map. The western area of the cold dust lane showing high column density exhibits colder dust emission (i.e., $T_d \sim$ 11-13 K) than the surroundings. The region near the massive star/cluster seems to have the highest temperature (i.e., $T_d \sim$ 19-20 K), and we can see a bubble of warm gas and dust surrounding the massive stars. The temperature map clearly shows the impact of massive star/s on the NGC 2316 cluster. It also suggests the formation of very young YSOs embedded in the cold and dense western lane.  
The panel (c) of the Figure \ref{massive} shows the color-composite image generated using the \emph{Spitzer} 4.5 $\mu$m, \emph{Spitzer} 3.6 $\mu$m, and \emph{UKIDSS} 2.2 $\mu$m band
images of the NGC 2316 region overlaid with the location of identified YSOs and the probable massive stars (red circles). We have overplotted the image with NVSS 1.4 KHz radio continuum emission contours. The \emph{Spitzer} 4.5 $\mu$m indicates the distribution of warm gas/dust, whereas \emph{Spitzer} 3.6 $\mu$m images include prominent PAH features at 3.3 $\mu$m, suggestive of photon dominant region (PDR) under the influence of feedback from the massive stars \citep[see e.g.][]{2004ApJ...613..986P}. 
In the panel (d) of Figure \ref{massive}, we show the \emph{Spitzer} ratio map of 4.5 $\mu$m/3.6 $\mu$m emission along with the radio contours and the location of the massive stars/cluster. This map's bright and dark regions trace the outflow activities from young stars and the PDRs, respectively (cf. \citealt{2017ApJ...834...22D}).  We can see the bright patches toward the north of the massive stars. Since these regions are also associated with younger YSOs, this bright patch suggests outflow activities from them.  Here, we also note that the bright patches are also visible in the western end of the dust lane, having young YSOs. The cluster/massive stars/ionized region is surrounded (mainly from the south direction) with a dark lane, suggesting the distribution of PDRs surrounding the massive stars. This kind of morphology is a typical feature of the H\,{\sc ii} region/MIR bubble created by the massive star/s. \citep{2023ApJ...953..145V,2008MNRAS.383.1241P,2007MNRAS.380.1141S}. 

\subsection{Molecular (CO) Morphology}
\label{section_MolecularMorphology}

\subsubsection{Channel Map}
\label{section_ChannelMap}

Figure \ref{fig_13CO_ChannelMap} shows the \tco\, channel map of the region. The entire range of emissions can be broadly divided into two regimes: the bluer regime in the range $\sim$[7.14, 10.63]\kms, and the redder regime in $\sim$[10.13, 15.11]\kms\, range. A SE-NW elongated structure can be seen at the bluer velocities, which then merge into a possible hub, most prominently seen in [8.63, 9.13] and [9.13, 9.63]\kms\, channels. At the redshifted velocity regime, a complex structure can be traced. Here, the main structure is also a prominent SE-NW elongated structure (most prominent in $\sim$[12.12, 12.62]\kms\, channel), to which various filaments seem to be joining. This redshifted structure also hosts our cluster of interest (see section \ref{group}). This redshifted elongated structure runs nearly parallel to the blueshifted one, and both seem to join in the velocity range $\sim$[10.13, 11.12]\kms. Hereafter, we refer to the blueshifted elongated structure as the Southern elongated structure and the redshifted one as the Northern elongated structure. The parallel elongated structures likely represent two different (filamentary) merging clouds. To examine this further, we look at the moment maps of the region in \co, \tco, and \cetno\, transitions in
Figure \ref{fig_CO_m012}.

\begin{figure*}
\centering
\includegraphics[width=\textwidth]{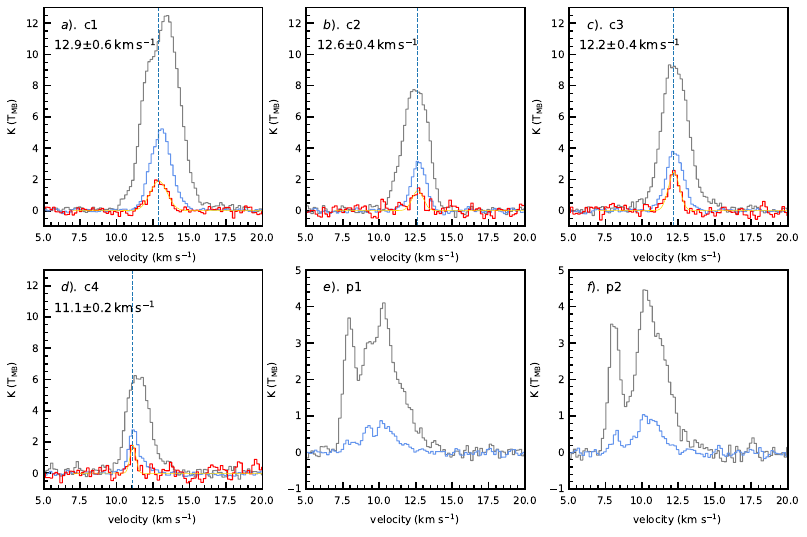}
\caption{
\co\, (grey), \tco\, (blue), and \cetno\, (red) spectra at the locations 
marked in Figure \ref{fig_CO_m012}(c), (g), and (h). 
The \cetno\, spectra (shown for c1-c4 only) have been scaled up by a factor 
of 3. 
For clumps c1-c4, the Gaussian fit to \cetno\, spectra, has been shown by 
yellow line, and the peak position is indicated by a dashed line. 
The peak location (velocity $\pm$ dispersion) is also specified on the images.
}
\label{fig_CO_spectra}
\end{figure*}

\begin{figure*}
\centering
\includegraphics[width=\textwidth]{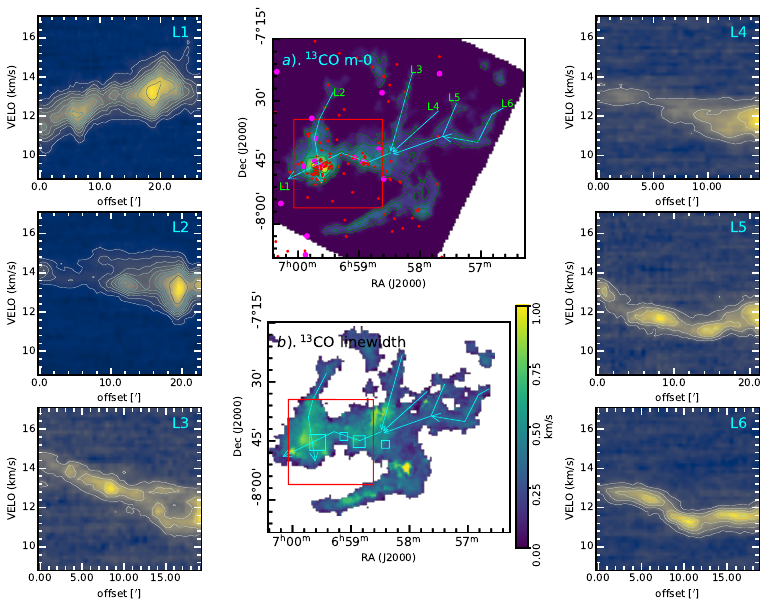}
\caption{
pv slices along the vectors (L1-L6) marked on the (a) \tco\, m-0 and (b) \tco\, linewidth images. The contour levels in the \tco\, m-0 map are at 1.2, 2.5, 4, 7, and 10 K\,\kms.  The four boxes on \tco\, linewidth image mark the locations of \cetno\, clumps (c1, c2, c3, and c4) from Figure \ref{fig_CO_m012}(c). The red box is our targeted region as shown in panel (a) of Figure \ref{3panel}. L1-L6 pv slices have contours marked at 0.45, 1, 1.5, 2, 2.5, 3, 3.5, 4, and 5 K. 
}
\label{fig_13CO_PV}
\end{figure*}

\subsubsection{Moment Maps}
\label{section_MomentMaps}

Figure \ref{fig_CO_m012} shows the emission regions for these transitions above 3$\sigma$ level ($\sigma$ being the noise level of respective cubes), i.e., we first identify clumps in the position-position-velocity (ppv) cubes 
which are above the requisite threshold level (3$\sigma$), and then collapse the cube in the appropriate velocity range to obtain moment-0 (m-0, Integrated Intensity), moment-1 (m-1, Intensity weighted velocity), and linewidth maps (see \citealt{2023ApJ...944..228M} for details). 
The most striking feature of the m-0 images (especially \co\, and \tco) of these 
regions is the web-like morphology, which depicts 
filamentary nature of the molecular cloud. 
Even in \co\, transition, which traces the cloud at a lower density \citep[$\sim$10$^2$\,cm$^{-3}$;][]{2019ApJS..240....9S}, such structures are prominent. \tco\, shows these features most prominently, whereas \cetno, as a tracer for the highest density regions, only shows a few clumps along the Northern elongated structure (see section \ref{section_ChannelMap}). In Figure \ref{fig_CO_m012}(c), these \cetno\, clumps have been labelled c1, c2, c3, and c4. The m-1 map (see Figure \ref{fig_CO_m012}(d) and (e)), shows the velocity field of the region, which reveals that the elongated structure in the south has a distinctly blueshifted velocity than that to its north. This is in consonance with the channel map discussed above. In the linewidth maps, two regions of interest, namely p1 and p2, have been marked on Figures \ref{fig_CO_m012}(g) and (h), respectively. The northern and southern elongated structures seem to interact at this location, i.e., roughly the junction where p1 and p2 are located. Both the \co\, (p1) and \tco\, (p2) velocity dispersion is high at this location. 

\subsubsection{Spectra}
\label{section_COspectra}

In this section, we focus on the spectra at the locations of \cetno\, clumps, namely c1-c4 labeled in Figure \ref{fig_CO_m012}(c), and locations of high-velocity dispersion, namely p1 and p2 marked in Figures \ref{fig_CO_m012}(g) and (h), respectively. In Figure \ref{fig_CO_spectra}, all three molecular spectra have been extracted for the \cetno\, clumps (c1-c4). As \cetno\, traces the densest parts, we fit a Gaussian to the \cetno\, spectra to obtain the velocity of the clumps. The Gaussian fit is shown as a yellow curve in Figures \ref{fig_CO_spectra}(a)-(d), with the fitting results of Gaussian peak position (i.e. velocity $\pm$ dispersion) mentioned on the respective images. The first thing that can be noticed here is that the clump velocities are redshifted as one moves from c4 to c1. The velocity dispersion is the largest for the clump c1 and smallest at c4. For c1, the optically thick \co\, spectra for c1 show a velocity profile that indicates self-absorption (and a red-asymmetrical peak), while the optically thinner \tco\, and \cetno\, profiles show a single peak \citep[see][]{LeungBrown_COSelfAbsorption_1977ApJ}. 

We next look at the molecular spectra at the locations of high dispersion in \co\, (Figure \ref{fig_CO_m012}(g)) and \tco\, (Figure \ref{fig_CO_m012}(h)), namely p1 and p2, respectively. Note that boxes p1 and p2 do not denote mutually exclusive locations. For these two locations, only \co\, and \tco\, spectra were noteworthy, and there was no discernible \cetno\, spectra. It can be seen in Figures \ref{fig_CO_spectra}(e) and (f), that \co\, spectra display two well-pronounced peaks, one at $\sim$7.5-8\kms\, and another at $\sim$10-10.5\kms. The relatively optically thinner \tco\, transition also shows two peaks at these velocities. This suggests that there are two cloud components at this location, further reinforcing the inference from the channel map (see section \ref{section_ChannelMap}) that the location p1/p2 represents the junction where the two disparate elongated filamentary structures are interacting. 

\subsubsection{Position-Velocity (pv) Maps}
\label{section_PVMaps}

Figure \ref{fig_13CO_PV} shows \tco\, m-0 and linewidth maps (from section \ref{section_MomentMaps}) on which six lines, labeled L1-L6 have been marked. pv slices extracted along all these directions using the \tco\, spectral cube have been shown in six respective side panels. The \tco\, cube was used for this analysis as it shows the filamentary nature of the cloud most prominently. We only consider the pv slices along directions associated with the northern elongated structure, since it is the one that hosts our cluster and since the southern elongated structure has been discussed to be a possible separate filament distinct from the northern one (see sections \ref{section_ChannelMap} and \ref{section_COspectra}). The first thing to notice in all the pv slices is a discernible velocity gradient. While for L1 to L4, the gradient is continuous from one end to another; for L5 and L6, the noticeable gradient is mainly up to about 11-12\arcmin\, offset. The morphology of the filaments seems to indicate a coalescence of filaments, wherein L6 merges into L5, and then L5, L4, and L3 merge into L1. Along the L1 vector line, are the c3, c2, and c1 clumps. Our cluster is associated with the c1 clump region (see section 3.1). The filament along the vector direction L2 also seems to be ending at the c1 clump. In the \tco\, linewidth map (cf. Figure \ref{fig_13CO_PV}(b)), the velocity dispersion seems to be increasing as one moves along the L1 vector, as also evidenced by increased velocity dispersion at c1 as opposed to c2 and c3 (see section \ref{section_COspectra} and Figure \ref{fig_CO_spectra}). The linewidth trend along the L2 vector is similar. This leads us to infer that there is a possible longitudinal flow of matter along the filaments towards the gravity well of the stellar cluster. 

\section{Discussion}\label{sec4}

We have tried to understand the formation history of the NGC 2316 cluster using the multi-wavelength (optical to Radio) data. We have used deep NIR data to study the stellar distribution in this region. We have found that the NGC 2316 cluster is a bit elongated (aspect ratio = 1.6) and small in size (0.4 pc). The stellar density map clearly shows the enhanced cluster region, but it also shows a dark lane with less dense stellar distribution, starting from the cluster and extending towards the western region. The foreground reddening in the cluster direction is $A_V$ = 1.55 mag, and there seem to be two probable massive stars (B2V-B1.5V) embedded in the region's nebulosity  ($A_V$ = 4 mag). The distance of this cluster was estimated from the proper motion data of the member stars as $1.3\pm0.3$ kpc. From the lifetime of the most massive stars in the cluster, we have put an upper age limit of this cluster as 12 Myr. As some of the member stars fall in the PMS stage in the CMD and some of the stars are showing excess IR emission (stars with discs around them, age $<$ 3 Myr \citealt{2009ApJS..181..321E}), we can safely conclude that this region is still forming young stars even-though the most massive star was born $\sim$12 Myr ago.

As this cluster has massive stars and very young low-mass stars, we have tried to study their formation and impact on the surrounding region using MIR/FIR/radio emission maps. The cluster is quite visible in MIR band images showing a distribution of warmed-up gas. Also, we can see a population of red stars in the region, which has a less dense stellar population. The FIR emission maps clearly show the distribution of cold gas/dust near the cluster and a dust lane from the cluster to the western direction. 
This dark lane corresponds to very low stellar densities. We have studied the location of YSOs in the region and found that most of these YSOs are located in the region of cold gas/dust. The YSO stellar density peaks near the cluster and at the western end of the dust lane. The YSO distribution is more extended than the cluster area. We have performed MST analysis on the YSOs location and identified two separate core/ARs in the region, i.e., CC/CAR and WC/WAR. The YSOs in the CC/CAR are found to be more massive than those inside the WC/WAR. Also, the YSOs in WC/WAR are more embedded than those in CC/CAR. As we know, the younger YSOs tend to be more embedded than the older ones, the WC/WAR seems to be hosting younger YSOs. We have derived the physical parameter of the identified cores/ARs of YSOs as listed in Table \ref{Tp1}. The CC/CAR is larger than WC/WAR. They both show elongated morphology and have similar stellar density and MST branch lengths.  The value of the Q parameter hints towards the hierarchical structure in these cores/ARs. The estimated CC/CAR cloud mass is 7-8 times that of WC/WAR. We have estimated the age and mass of the YSOs in these core/ARs from their SED (\citealt{2017MNRAS.467.2943S}) and found that the CC/CAR YSOs are older than the WC/WAR YSOs. Also, CC/CAR YSOs are much more massive than those inside the WC/WAR. The star formation efficiency (SFE) calculations hint toward a slightly higher SFE towards WC/WAR.

We have found the probable massive stars located at the peak of the radio continuum emission near the peak of the YSO density distribution. The massive stars are inside the bubble structure, as the MIR images show. No radio emission is detected in the western end of the dust lane.  The column density maps from the $Hershel$ also show the long dust lane with a peak at the western end. Most of the YSOs are located in the region of higher column densities. The outflow activities from the young YSOs are also visible near the cluster and at the western end of the dust lane. The temperature shows a clear warmed-up region near the cluster, whereas the western region seems to have a much lower temperature. Thus, an interaction might occur between the massive star and the surrounding material. The $Spitzer$ ratio maps also delineate the bubble-shaped PDR surrounding the massive stars. The Massive stars affect their surroundings through feedback pressure, which plays a crucial role in the self-regulation of star formation. For B2V-B1.5V type stars with a projected distance of $D_s \sim$ 0.7 pc between the Class I YSOs and the massive star, we have calculated the total pressure value as $\sim10^{-10}$ dynes\,cm$^{-2}$ considering Lyman continuum photons, mass-loss rate, wind velocity, and bolometric luminosity as $7.76 \times 10^{44}$ photons\,s$^{-1}$ \citep{1973AJ.....78..929P},  $1.99 \times 10^{-10}$ M$_{\odot}$ yr$^{-1}$ \citep{10.1111/j.1365-2966.2011.19143.x,2016ApJ...833...85B}, 700 \kms \citep{10.1111/j.1365-2966.2011.19143.x,2016ApJ...833...85B}, and 5011.87 L$_{\odot}$ \citep{1973AJ.....78..929P}, respectively for B2.0V type star; and $1.0 \times 10^{46}$ photons\,s$^{-1}$ \citep{2004A&A...427..839C}, $1.02 \times 10^{-8}$ M$_{\odot}$ yr$^{-1}$ \citep{2023A&A...678A.172P}, 1200 \kms \citep{2023A&A...678A.172P} and 9549.93 L$_{\odot}$ \citep{2005IAUS..227..389C}, respectively, for B1.5V type star (for more details on pressure calculation, please refer \citealt{2023ApJ...953..145V}). The calculated total pressure is greater than the internal pressure ($\sim$$10^{-11}$ - $10^{-12}$ dynes\,cm$^{-2}$) of a typical molecular cloud for particle density $\sim$$10^{3}$ - $10^{4}$ cm$^{-3}$ at a temperature of $\sim$20 K (refer Table 2.3 of \citealt{1980pim..book.....D}) and is sufficient to trigger a collapse of the molecular clouds to form young stars.


We have studied the molecular morphology in the region using high-resolution molecular (\co, \tco, \cetno) data from the PMO survey. From the channel maps of \tco, we have found two filamentary clouds, at separate velocities, interacting in this region. The northern elongated structure/cloud is more filamentous than the southern cloud. Our cluster, along with several filaments, resides in the northern clouds at 12.12-12.62 \kms. We have also studied m-0 (Integrated Intensity), m-1 (Intensity-weighted velocity), and linewidth maps at all three CO transitions. We can see the web-like filamentary structure in the m-0 maps.  The \cetno m-0 map (tracer for the highest density regions) only shows a few clumps along the Northern elongated structure (c1, c2, c3, and c4) mainly located on our region's dust lane/cluster. The m-1 map clearly shows that the elongated structure in the south has a distinctly blueshifted velocity compared to its north. The linewidth maps indicate the region (p1/p2) where the northern and southern elongated structures seem to interact (velocity dispersion is high at this location). We have studied the velocity spectra at this interacting junction (p1/p2) as well as at the four identified dense (\cetno) cores (c1, c2, c3, c4) in the molecular transitions. The clump velocities are redshifted as one moves from c4 to c1. The velocity dispersion is largest for the clump c1 and smallest at c4. At c1, we found the signature of infall of molecular material. Both p1 and p2 regions' spectra have peaks at 7.5-8\kms\, and 10-10.5\kms, in both \co and \tco\, transitions, suggesting two cloud components at this location. Thus, p1/p2 represents the junction where the two disparate elongated filamentary structures interact and seem to be a good candidate for filament-filament interaction studies in the future. 

We have identified six filamentary structures (L1-L6) in the northern elongated structure/cloud using the \tco\, maps. We then extracted the pv slices along these directions and found they all show a discernible velocity gradient. The morphology of the filaments seems to indicate a coalescence of filaments, wherein L6 merges into L5, and then L5, L4, and L3 merge into L1. Along the L1 vector line are the c3, c2, and c1 clumps. Our cluster is associated with the c1 clump region. The filament along the vector direction L2 also seems to be ending at the c1 clump. In the \tco\, linewidth map, the velocity dispersion seems to increase as one moves from the west to towards the cluster center. All these results lead us to infer that there is a possible longitudinal flow of matter along the filaments towards the gravity well of the stellar cluster.

Recent observational investigations have validated the association of HFS with cluster formation (for example, \citealt{2012A&A...540L..11S,2019A&A...631A...3M,2022A&A...658A.114K,2022A&A...660L...4B,2023ApJ...944..228M,2023MNRAS.tmp.3249D,2023JApA...44...23D,2023ApJ...958...51D,2023ApJ...958...17P}). Recently, \citet{2020A&A...642A..87K} reported a broad evolutionary sequence of the HFS describing the formation framework of the massive stars from molecular gas in four consecutive stages. In the first stage, the dense filaments approach each other, setting up the preliminary conditions for forming HFS. The formation of HFS can be initiated by several processes, such as energetic stellar wind bubbles, flow-driven filaments,  expanding ionization fronts, supernova shocks, etc. In the second stage, filaments merge, forming a `hub'. They merge in such a manner that a twist at the overlapping zone causes the flattening of the hub. In the third stage, the initial shock followed by self-gravity causes the amplification in the density of the hub. This results in a longitudinal flow towards the hub that, in turn, forms massive stars. Eventually, in the fourth stage, the ionization feedback and the radiation from the massive star configure the filaments as pillars and leave a mass-segregated embedded cluster at the hub.

While the main cluster region seems to be an HFS between the third and fourth stages, in which the NGC 2316 cluster is probably at the hub, and the dark lane is the main filamentary structure; the location p1/p2 seems to be at the second stage, wherein filament-filament interaction and possible merger to form a hub can take place. The structure associated with the NGC 2316 cluster is divided into many other filamentary structures. Almost all the star formation in the region is happening in this HFS. The age analysis of YSOs/massive stars suggests that the star formation first started in the NGC 2316 region (gravity well), and subsequently, it went on to the other filament nodes.

\section{Conclusion}\label{sec5}

We present a multi-wavelength analysis of the NGC 2316 star cluster and its surroundings using the data from various telescopes/archives. This region shows evidence of recent star formation and is associated with molecular filamentary structures showing signatures of gas infall. The following conclusions are made from our study:

\begin{enumerate}
      
\item
We studied the stellar density distribution in this region using a deep NIR catalog and found stellar clustering (NGC 2316) along with a lane having less stellar distribution towards the west direction. The NGC 2316 cluster is elongated in shape (aspect ratio = 1.6) and small in size (R$_{cluster}$ = 0.4 pc). We also identified 29 optical members of this cluster using the \textit{Gaia} DR3 PM data. The distance of this cluster is estimated as $1.3\pm0.3$ kpc.
 
\item We estimated the minimum reddening in the direction of this cluster as $A_V$ = 1.55 mag. We have also found a couple of massive stars (B2.0V-B1.5V, age $\sim$12 Myr) embedded ($A_V$ = 4 mag) inside this cluster. As this cluster also hosts YSOs (age $<$ 3 Myr), we can safely conclude that this region still forms young stars even though the most massive star was born $\sim$12 Myr ago.  

\item 
We identified a cold gas/dust lane from the cluster towards the western direction through FIR maps.  It is the same lane evident in the stellar distribution. This dust lane hosts a population of red stars embedded in its nebulosity. The YSOs in the region mostly follow the cold gas/dust distribution and their number density peaks near the NGC 2316 cluster and at the western end of the dust lane. The regions near the NGC 2316 cluster show a warmed-up gas/dust distribution, suggesting stellar feedback from the massive stars.

\item
Through MST analysis, we identified two separate core/ARs in the region, i.e., CC/CAR and WC/WAR, at both ends of the identified dust lane. The physical properties of these cores/ARs are reported in this study. The CC/CAR coincides with the NGC 2316 cluster and is more massive than WC/WAR at the dust lane's western end. The YSOs in the CC/CAR are also found to be more massive than those inside the WC/WAR. The WC/WAR hosts younger YSOs and has slightly better SFE than CC/CAR.

\item 
The MIR emission/temperature maps and distribution of PDR in the region suggest a bubble of warmed-up gas near the massive stars/NGC 2316 cluster. The location of massive stars also coincides with the peak of the radio continuum emission. Towards the north of the massive stars, we can see the distribution of younger YSOs behind the bubble. This morphology resembles the H\,{\sc ii} region. The total pressure value ($1.53 \times 10^{-9}$  dynes\,cm$^{-2}$) from massive stars is sufficient enough to trigger star formation in their surrounding molecular clouds.

\item
 The column density map shows a long gas/dust lane with a peak in the western direction of this region. The YSOs are located in the region of higher column densities. The western end of the gas/dust lane seems to have a much lower temperature and has no radio emission. The outflow activities from the young YSOs are visible near the cluster and at the western end of the dust lane, suggesting recent star formation. The western end of the gas/dust lane seems to favor low-mass star formation, whereas the cluster's end favors bit-massive star formation. There is also a sequence in the star formation where the cluster center started earlier than the western end.

\item 
We studied this region's molecular cloud morphology/dynamics using the high-resolution molecular (\co, \tco, \cetno) maps. We found two elongated structures/molecular clouds interacting with each other. The northern elongated structure, which hosts the presently studied region, seems to have many filamentary structures. We have identified four dense clumps in this structure, and their velocities are found to be redshifted as one moves toward the cluster. The velocity dispersion is also the largest for the clump near the cluster, and it gets smaller as we go away from the cluster.  We have also identified six filamentary structures in the northern elongated structure/cloud. The morphology of the filaments and the pv maps indicate a coalescence of filaments and a possible longitudinal flow of matter along the filaments towards the gravity well of the stellar cluster.

\end{enumerate}

Finally, this entire region seems to be an HFS, in which the NGC 2316 cluster is probably the hub and the dark lane is the main filamentary structure along with many other filamentary sub-structures. The star formation might have started in the NGC 2316 region first (gravity well), and subsequently, it went on to the other filament nodes.

\section*{Acknowledgements}
The observations reported in this paper were obtained by using the 1.0m ST and 3.6m DOT telescopes at ARIES, Nainital, India. This research used data from the Milky Way Imaging Scroll Painting (MWISP) project, a multi-line survey in 12CO/13CO/C18O along the northern galactic plane with PMO-13.7\,m telescope. We are grateful to all the members of the MWISP working group, particularly the staff members at PMO-13.7\,m telescope, for their long-term support. MWISP was sponsored by the National Key R\&D Program of China with grant 2017YFA0402701 and by CAS Key Research Program of Frontier Sciences with grant QYZDJ-SSW-SLH047.
This publication uses data from the Two Micron All Sky Survey, a joint project of the University of Massachusetts and the Infrared Processing and Analysis Center/California Institute of Technology, funded by the National Aeronautics and Space Administration and the National Science Foundation. This work is based on observations made with the \emph{Spitzer} Space Telescope, operated by the Jet Propulsion Laboratory, California Institute of Technology, under a contract with the National Aeronautics and Space Administration. This publication uses data products from the Wide-field Infrared Survey Explorer, a joint project of the University of California, Los Angeles, and the Jet Propulsion Laboratory/California Institute of Technology, funded by the National Aeronautics and Space Administration. DKO acknowledges the support of the Department of Atomic Energy, Government of India, under Project Identification No. RTI 4002. AV acknowledges the financial support of DST-INSPIRE (No.$\colon$ DST/INSPIRE Fellowship/2019/IF190550).

\vspace{5mm}




\appendix

\section{Membership analysis of the stars in the NGC2316 complex} \label{memberg}

\begin{figure*}
        \centering
        \includegraphics[width=0.48\textwidth]{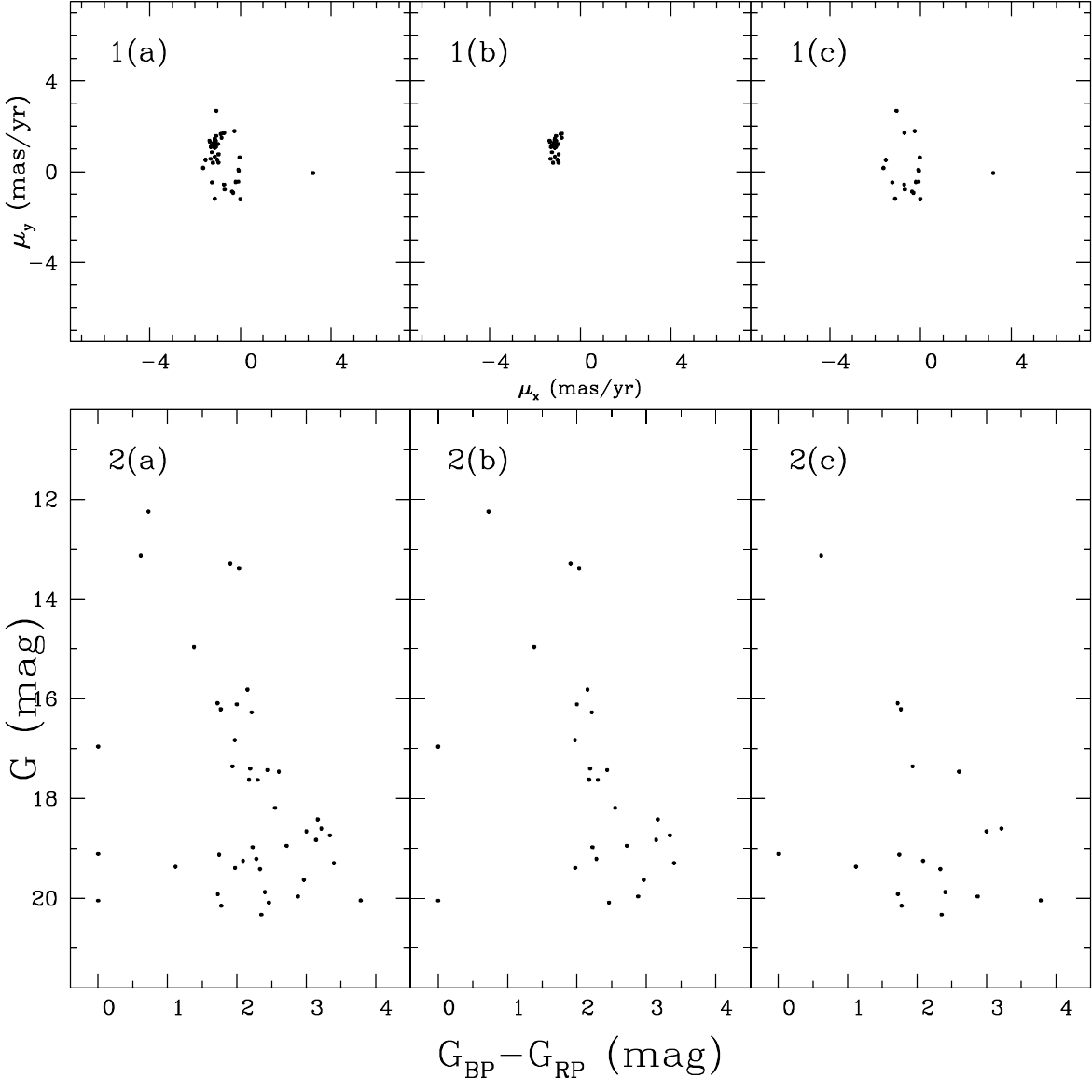}
        \includegraphics[width=0.48\textwidth]{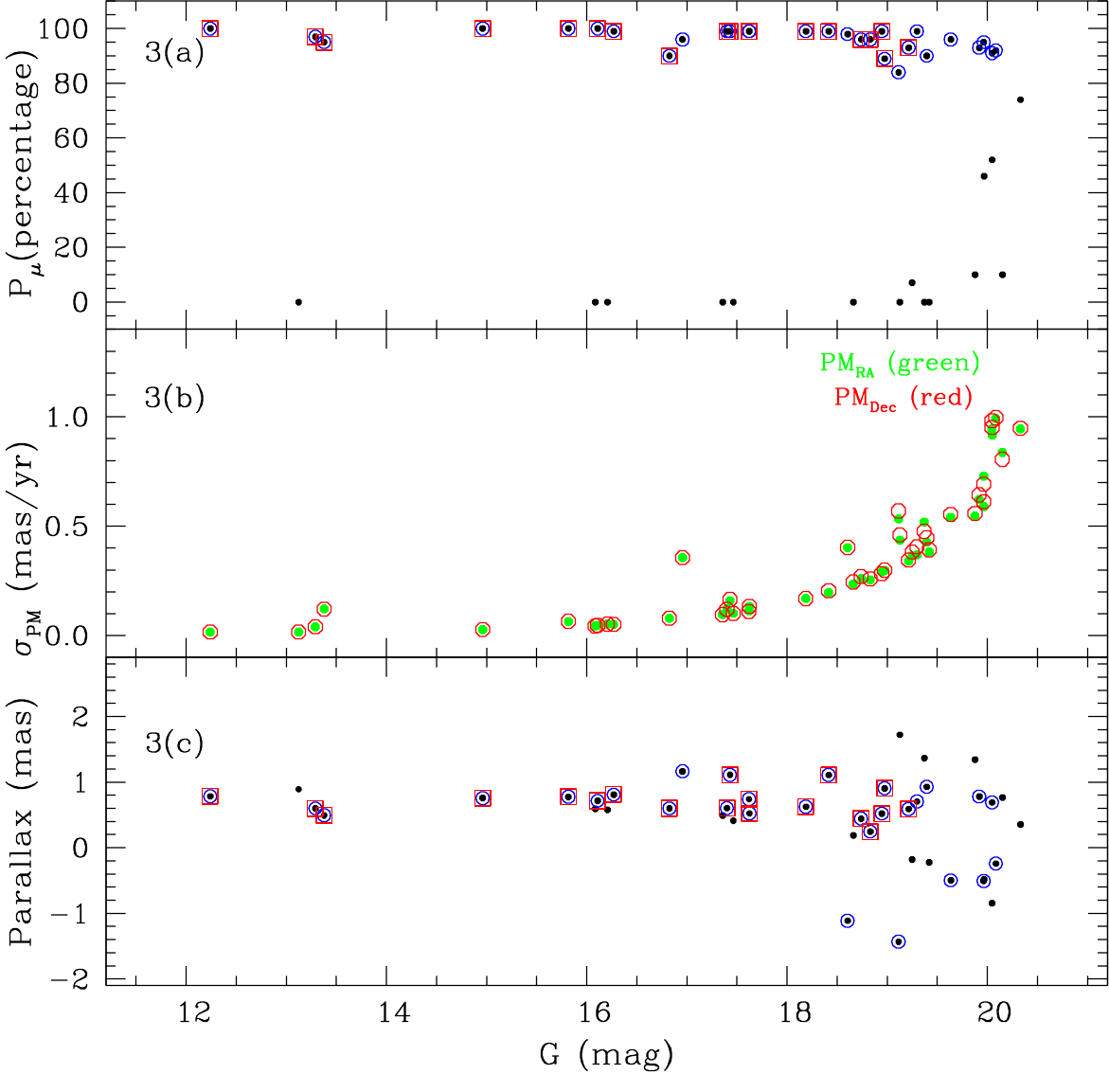}
                \caption{\label{pm1} Panel 1 \& 2: PM vector-point diagrams (panel 1) and
        {\it Gaia} DR3 $G$ vs. $(G_{BP} - G_{RP})$ CMDs (panel 2) for stars located in the NGC 2316 cluster region.
        The left sub-panels (1(a) and 2(a)) show all-stars, while the middle (1(b) and 2(b)) and the right sub-panels (1(c) and 2(c)) show
        the probable cluster members and field stars, respectively.
        Panel 3: Membership probability P$_\mu$, PM errors $\sigma_{PM}$ and parallax of stars as
        a function of $G$ magnitude for stars in the NGC 2316 cluster region.
        Stars with P$_\mu \geq$ 80 are considered members of the NGC 2316 cluster and are shown by blue circles. Square box symbols represent the stars with PM errors $\leq$ 0.4 mas (considered for estimating the distance of the NGC 2316 cluster).
        }
        \end{figure*}

        \textit{Gaia} data located within the hull (cluster region) and having PM error $\leq$ 1 mas\,yr$^{-1}$, has been used to determine their membership probability for the cluster. 
        PMs $\mu_{x}$ and $\mu_{y}$  are plotted as vector-point diagrams (VPDs) in panel 1 of
        Fig. \ref{pm1}. The panel 2 shows the corresponding $G$ versus $G_{BP} - G_{RP}$
        color-magnitude diagrams (CMDs).
        The dots in sub-panel 1(a) represent the PM distribution of all the stars
        where a prominent  clump within a radius of $\sim$1 mas yr$^{-1}$ centered at
        $\mu_{xc}$ = -2.456 mas yr$^{-1}$, $\mu_{yc}$ = +1.243 mas yr$^{-1}$ can bee seen.
        This population of stars has almost similar PMs and has a high probability of cluster membership.
        The remaining stars with scattered PM values are most probably a field population.
        This is clearer in the VPDs and CMDs of probable cluster and field populations, as shown in the
        sub-panels 1(b), 1(c), 2(b), and 2(c), respectively.
        The probable cluster members show well-defined MS in the CMD, usually seen for a similar population of stars.
        On the other hand, the probable field stars are quite obvious by their broad distribution in the CMD.
        Assuming a distance of 1.1 kpc  \citep{1994LNP...439..175F} and a
        radial velocity dispersion of 1 km s$^{-1}$ for open clusters \citep{1989AJ.....98..227G},
        the expected dispersion ($\sigma_c$) in PMs would be $\sim$0.2 mas yr$^{-1}$.
        For field stars we have: $\mu_{xf}$ = 0.218 mas yr$^{-1}$, $\mu_{yf}$ = $-$0.380 mas yr$^{-1}$, $\sigma_{xf}$ = 4.60 mas yr$^{-1}$ and $\sigma_{yf}$ = 1.81 mas yr$^{-1}$.
        By using these values, we have calculated the frequency distribution of cluster stars ($\phi^{\nu}_c$) and field stars ($\phi^{\nu}_f$), as determined in \citet{2020MNRAS.498.2309S}.
        The membership probability (ratio of the distribution of cluster stars with all the stars) for the $i^{th}$ star is:

        \begin{equation}
        P_\mu(i) = {{n_c\times\phi^\nu_c(i)}\over{n_c\times\phi^\nu_c(i)+n_f\times\phi^\nu_f(i)}}
        \end{equation}

        where $n_c$ (= 0.55) and $n_f$(= 0.45) are the normalized number of stars for the cluster
        and field regions ($n_c$+$n_f$ = 1).

        The membership probability estimated as above, errors in the PM,
        and parallax values are plotted as a function of $G$ magnitude in
        panel 3 of Fig. \ref{pm1}. As can be seen in this plot, a high
        membership probability (P$_\mu \geq$ 80 \%) extends down to $G\sim$20 mag.
        At brighter magnitudes, there is a clear separation between
        cluster members and field stars, supporting the effectiveness of this
        technique. PM errors become very high at faint limits, and the
        maximum probability gradually decreases at those levels. Except
        for a few outliers, most stars with high membership
        probability (P$_\mu \geq$ 80 \%) follow a tight distribution.
        Finally, based on the above analysis,
        29 stars were assigned as cluster members based on their high membership probability P$_\mu$ ($\geq$ 80 \%; blue circles in Fig. \ref{pm1} (panels 3a and 3c)).

\bibliography{thebibliography}{}
\bibliographystyle{aasjournal}

\end{document}